\documentclass[11pt]{article}
\usepackage[T1]{fontenc}
\usepackage{lmodern,microtype}
\usepackage[margin=1in]{geometry}
\usepackage{amsmath,amssymb,graphicx,booktabs}
\usepackage[mathlines]{lineno}
\usepackage[super,sort&compress]{natbib}

\usepackage[dvipsnames]{xcolor}
\usepackage[colorlinks=true,allcolors=MidnightBlue]{hyperref}

\usepackage{caption}
\DeclareCaptionFont{linenumbers}{\internallinenumbers}
\newcommand{\HC}{\mathrm{HC}}

\title{Similar visual tuning conceals different recovery dynamics in a model of macaque V1}
\author{\parbox{\dimexpr\textwidth-2\tabcolsep\relax}{\centering
\normalsize Xialei Huang$^{1}$, Qingyu Yan$^{2}$, Zheng Bian$^{3,4,*}$, and
Zhuo-Cheng Xiao$^{1,5,6,*}$\\[0.8em]
\small
$^{1}$Division of Art and Science, NYU Shanghai, Shanghai 200124, China\\[0.3em]
$^{2}$Applied Physics Graduate Program, Rice University, Houston, TX 77005, USA\\[0.3em]
$^{3}$ Instituto de Matemática, Estatística e Ciência da Computação da Universidade de São Paulo, São Paulo, SP 05508-090, Brazil\\[0.3em]
$^{4}$Department of Mathematics, Imperial College London, London SW7 2AZ, UK\\[0.3em]
$^{5}$NYU-ECNU Institute of Mathematical Sciences at NYU Shanghai,
Shanghai 200124, China\\[0.3em]
$^{6}$NYU-ECNU Institute of Brain and Cognitive Science at NYU Shanghai,
Shanghai 200124, China\\[0.5em]
$^{*}$Corresponding authors
}}
\date{}
\begin{document}
\maketitle

\begin{abstract}
Biologically detailed neuronal networks can model how brain circuits perform observed functions, but their high computational cost limits the comparisons needed to test biological hypotheses. To address this limitation, we extend a multiscale approach in which responses of small populations of spiking neurons are computed in advance and reused across a larger cortical circuit. In particular, we approximate these local responses by deep neural networks and achieve an estimated acceleration of the order $10^5$ for steady-state evaluation relative to spiking network simulation. Beyond computational acceleration, the smooth deep surrogate of local responses provides derivatives, allowing us to analyze linear stability and to trace how a perturbation in brain activity is transmitted through different pathways. We demonstrate the approach in a model of an input layer of macaque primary visual cortex (V1), testing circuits under visual stimulation and controlled perturbations. We resolve pathway contributions to stability, strong recurrent responses, and slow recovery. Yet different combinations of pathway strengths can support similar responses to visual stimuli while producing different recovery dynamics. Therefore, Steady responses and perturbation recovery place complementary constraints on candidate circuits. Overall, our approach offers efficient evaluation with an interpretable circuit structure, making systematic comparisons across stimuli and biological hypotheses practical.
\end{abstract}

\section*{Introduction}

Multiscale modeling offers an efficient and interpretable way to construct and simulate biologically detailed models of cortical circuits, which are otherwise often too computationally demanding. The idea is to treat neuronal activity at different anatomical scales separately, using the results at one scale to inform computations at another. In a cortical layer with spatially homogeneous local circuitry, this means first computing the response of a small local neuronal circuit to input signals, which are reused for all other populations. Then, one can combine these responses through spatial connections across the cortical layer. The resulting coarse-grained model retains biologically relevant features, including cell types, neuronal responses, and spatial connectivity, while reducing the cost of simulation. Therefore, multiscale cortical modeling can extend beyond a small piece of cortex, allowing the simultaneous examination of multiple brain functions a larger range of circuit configurations\citep{Xiao2024,Xiao2026}.

The macaque visual system has proven to be a fruitful example to illustrate these ideas. Along the magnocellular pathway, signals relayed through the lateral geniculate nucleus (LGN) reach layer 4C$\alpha$ of primary visual cortex (V1)\citep{BlasdelFitzpatrick1984}. V1 neurons prefer different stimulus orientations: some respond strongly to one angle and weakly to another, whereas others are broadly tuned\citep{Ringach2002}. This diversity is present within layer 4C, including its magnocellular recipient sublayer\citep{Ringach2002}. Layer 4C$\alpha$ neurons are also sensitive to contrast in visual simuli\citep{BlasdelFitzpatrick1984}, and recordings reveal spatial- and temporal-frequency tuning and direction selectivity\citep{Hawken1988,chariker2022computational}. Thus, several visual response properties that can constrain cortical models are already measurable at this early cortical stage.

Biologically detailed network models of layer 4C$\alpha$ reproduce these visual functions simultaneously, using a single set of equations and parameters across the stimuli tested\citep{Chariker2016,Chariker2020,chariker2022computational}. These models describe interactions among LGN input, excitation and inhibition within layer 4C$\alpha$, and feedback from layer 6. Based on this, Xiao et al.\ developed the multiscale construction, leading to a comprehensive and scalable model of layer 4C$\alpha$\citep{Xiao2024}. They later used monocular and binocular responses to constrain cross-column circuitry, and used it to investigate how signals from the two eyes first interact in cortex\citep{Xiao2026}. Together, this line of work demonstrated how a multiscale cortical model constrained by known anatomy and physiology could efficiently generate and test biological hypotheses about circuitry that was less well understood.

However, there is one crucial limitation. The previous works mentioned above rely heavily on firing-rate maps representing time-averaged activity in a nonequilibrium steady state (NESS): a driven state with steady population statistics despite continuing neuronal activity. Although the multiscale method computes the firing-rate maps efficiently, a steady-response calculation alone does not establish whether all nearby activity patterns return to NESS. Since cortical activity continues to fluctuate during repeated presentations of the same stimulus\citep{Churchland2010}, small deviations from an unstable firing-rate map can grow rather than subside. These considerations motivate us to examine the stability of NESS.


Moreover, even when a NESS is stable, recovery from perturbations can be nontrivial. Recurrent interactions can temporarily amplify some perturbation patterns, while slowly decaying patterns shape the later stages of recovery\citep{MurphyMiller2009,Hennequin2012}. In fact, recovery dynamics offers a complementary constraint on the computational model.  Studies of other neuronal circuits show that different parameter combinations can generate similar ongoing activity\citep{Prinz2004}, yet experiments showed that two circuits producing similar dynamics can respond differently to the same perturbation\citep{Powell2021}. 
In V1, coordinated changes in recurrent excitation, inhibition, and feedback from other layers may approximately preserve a steady firing-rate map while substantially changing transient responses. Comparing both NESS and recovery dynamics could therefore distinguish pathway combinations that NESS alone leave unresolved.

Indirect experimental evidence for pathway-dependent recovery has been found in mouse V1, where cortical activity rapidly decayed after thalamic silencing and decayed more slowly when cortical inhibition was reduced under anesthesia\citep{Reinhold2015}. In macaque, which pathways can shape the amplification and persistence, and how do changes in the pathway connectivity strengths move the circuit toward or away from instability? 
To answer these questions, we perform perturbation analysis near NESS. Compared to costly animal experiments, which usually require simultaneous recordings across multiple areas\citep{CasteloBranco1998,Bastos2014,Han2021}, our computational approach can be implemented in an efficient, biologically realistic circuit model. 

Motivated by the above considerations on NESS stability and recovery dynamics, we propose in the current work to use deep neural networks (DNNs) to represent the local response functions in the previous multiscale approach. The upshot is two-fold. Evaluating an explicit learned function is much faster than looking up and interpolating in a large response library as in previous work, further accelerating steady-response calculations and parameter comparisons. Also, we impose that the learned functions be smooth and hence provide derivatives: they describe how small changes in a population's inputs change its firing rate. Combined with the spatial connections among populations, these derivatives form the local response matrix, or Jacobian, which describes how a small perturbation is transmitted across the whole circuit. 

Using these learned response functions, we derive a population-rate equation and analyze its linearized dynamics near NESS to assess stability, transient amplification, and slow recovery. We identify how distinct circuit pathways shape these dynamics and how the resulting population activity affects an orientation readout. We then use fixed-point-preserving interventions to compare pathway contributions at a common NESS, distinguishing pathway combinations that produce the same steady response. Together, these comparisons show how coordinated pathway changes can preserve broad visual tuning while changing recovery dynamics.


\section*{Results}

\subsection{A multiscale V1 model accelerated and made differentiable by deep surrogates}

\subsubsection{Layer 4C\texorpdfstring{$\alpha$}{alpha} and its spatial organization}

Layer 4C$\alpha$ (L4 hereafter) is an input layer of macaque V1 along the magnocellular visual pathway\citep{BlasdelFitzpatrick1984}. We represent it as a sheet of cortex divided into hypercolumns (HCs), following the organization of the parent models\citep{Chariker2016,Xiao2024,Xiao2026}. Each model HC contains $\sim4000$ cells. Neurons within an HC are assigned approximately the same receptive-field location, meaning they respond to stimuli in approximately the same part of the visual field. They differ, however, in the orientations to which they respond most strongly.

Optical maps of macaque V1 show orientation domains (ODs), including regions where preferred angles progress around pinwheel centers\citep{ObermayerBlasdel1993}. Our model uses an idealized, regularly repeated arrangement of these domains with periodic boundaries(Fig.~\ref{fig:1}A). The short bars indicate the orientations favored by the assigned LGN afferent configurations. This spatial organization provides the setting for orientation selectivity (OS), the ability of a neuron or population to respond differently to gratings at different angles. The interaction generated by the full circuit determine its actual response to visual stimuli. In particular, recurrent interactions lead to orientation preferences between discrete angles assigned to ODs\citep{Chariker2016,Xiao2024}.

\vspace{0.1in}
\noindent\textbf{Input pathway decomposition.} For each cell on L4, we distinguish feedforward input from LGN, recurrent excitation and inhibition within L4, and feedback from layer 6 (L6). The magnocellular LGN input follows the anatomically constrained, sparse-afferent setup of Chariker et al.\citep{ConnollyVanEssen1984,Chariker2016}. In the model, simple cells receive spatially aligned ON and OFF LGN afferents depending on the OD in which it lies. These ON and OFF LGN cells respond to increases and decreases in luminance, respectively. Their combined spike timing depends on the orientation of the grating relative to that alignment. Complex cells are assigned less direct LGN input and stronger dependence on cortical drive\citep{Chariker2016,Xiao2026}.

Recurrent excitation and inhibition are determined by the network structure and activities of cells nearby. Anatomical tracing in macaque V1 identifies projections from L4 to L6 and substantial projections from L6 back to L4\citep{Fitzpatrick1985}. Following Chariker et al.\, we represent the returning L6 input by a prescribed rule whose amplitude depends on L4 excitatory activity\citep{Chariker2020,Xiao2026}. This effective rule reduces the more complex L6 circuitry to a feedback pathway within the L4 model. Therefore, a change in L4 activity changes the input returned through that pathway. 

\subsubsection{Overview of the multiscale approach}

\noindent\textbf{Local dynamics.} We divide each HC into small groups of nearby neurons, called \textit{local populations} or \textit{pixels}. Each pixel contains three different types of cells: simple excitatory cells (S), complex excitatory cells (C), and inhibitory cells (I). Simple and complex cells are kept separate due to their different relative contributions of thalamic and cortical input. Neurons in a local population respond to inputs from outside the population as well as to interactions within it. We refer to their neuronal activity on this scale as local dynamics. Because the local architecture is similar across the modeled layer, the same response calculation can be used at different locations, with the inputs varied to reflect the surroundings. We describe this input--response relationship by three local response functions, $\Psi^S$, $\Psi^C$, and $\Psi^I$, which map LGN, recurrent excitatory (L4E), recurrent inhibitory (L4I), and L6 inputs to the corresponding mean firing rates (Methods, Eq.~\ref{local_response_function_explicit}), that is,
\[
f^S_p = \Psi^S(\text{LGN input to } p, \text{ L6 input to } p, \text{ L4E input to } p, \text{ L4I input to } p),
\]
where $f^S_p$ represents the mean firing rate of S cells in the $p$th pixel, with analogous definitions for C and I cells.

Our working hypothesis is that neurons of the same type within a pixel receive statistically equivalent inputs and are represented by a common mean response. Local populations are chosen small enough to ensure this hypothesis, while grouping enough neurons to reduce the computational cost. Throughout this paper, we choose to group 40 neurons in a pixel, though this number can change depending on the desired resolution of pixels (see Fig.~S4). 

In practice, the LGN inputs are precomputed and supplied to the local response calculations. We first assign a couple of spatial configurations of LGN receptive fields, then pool the summed LGN spiking series from simulations of these LGN cells when driven by a visual stimuli\citep{Xiao2024,Xiao2026}.  On the other hand, L4E, L4I, and L6 inputs are set as Poisson series parametrized by their rates, which are determined by the interactions among pixels (see below).

\vspace{0.1in}
\noindent\textbf{Coarse-grained dynamics.} The other computations are for interactions between local populations and determine the L4 and L6 input rates (for type-Q cells in pixel $p$, that is, $n^{Q\rm L6}_p$, $n^{Q\rm L4E}_p$, and $n^{Q\rm L4I}_p$). These coarse-grained (CG) interactions are specified by spatial connectivity kernels derived from the cell types, connection probabilities, and spatial ranges in the biological model. $C_{(p,Q)\leftarrow(p',R)}$ is the average number of type-$R$ cells in pixel $p'$ presynaptic to a type-$Q$ cell in pixel $p$, where $Q,R\in\{S,C,I\}$. Therefore, given the firing rates of the source populations $p'$, the kernels determine the inputs received at each target population $p$ (Methods, Eq.~\ref{local_response_function_explicit}). The local response functions $\Psi$ then supplies new rates for those inputs. Thus information passes between the two scales: population activity determines input, and local neuronal responses determine the population activity as the output, evolving the whole circuit for one step on time.

\vspace{0.1in}
\noindent\textbf{Pixel firing rates.} At each pixel $p$, the CG state consists of three mean rates, $(f^S_{p},f^C_{p},f^I_{p})$. The main field contains four by four HCs, each subdivided into ten by ten pixels, giving 4,800 rate variables. In activity maps, we display either the excitatory rate, combining S and C according to their population proportions, or the inhibitory rate. A steady cortical response is a self-consistent collection of these rates: the inputs generated by the field sustain the local rates from which they were computed. At the neuronal level, this is a driven, continually active state with steady population statistics, or a nonequilibrium steady state.

\subsubsection{Preservation of modeled visual functions}
\label{Sect:1.3}

Following the tests in Chariker et al.\citep{Chariker2020}, we focus on orientation selectivity and contrast response. These responses provide the visual setting for the pathway and perturbation analyses in the rest of the paper. We therefore use drifting gratings for all visual stimulus. The drifting grating always fixes a spatial frequency of $2.5$ cycles/degree and a temporal frequency of $10$ Hz.

\begin{figure}[p]
\centering\includegraphics[width=0.99\textwidth]{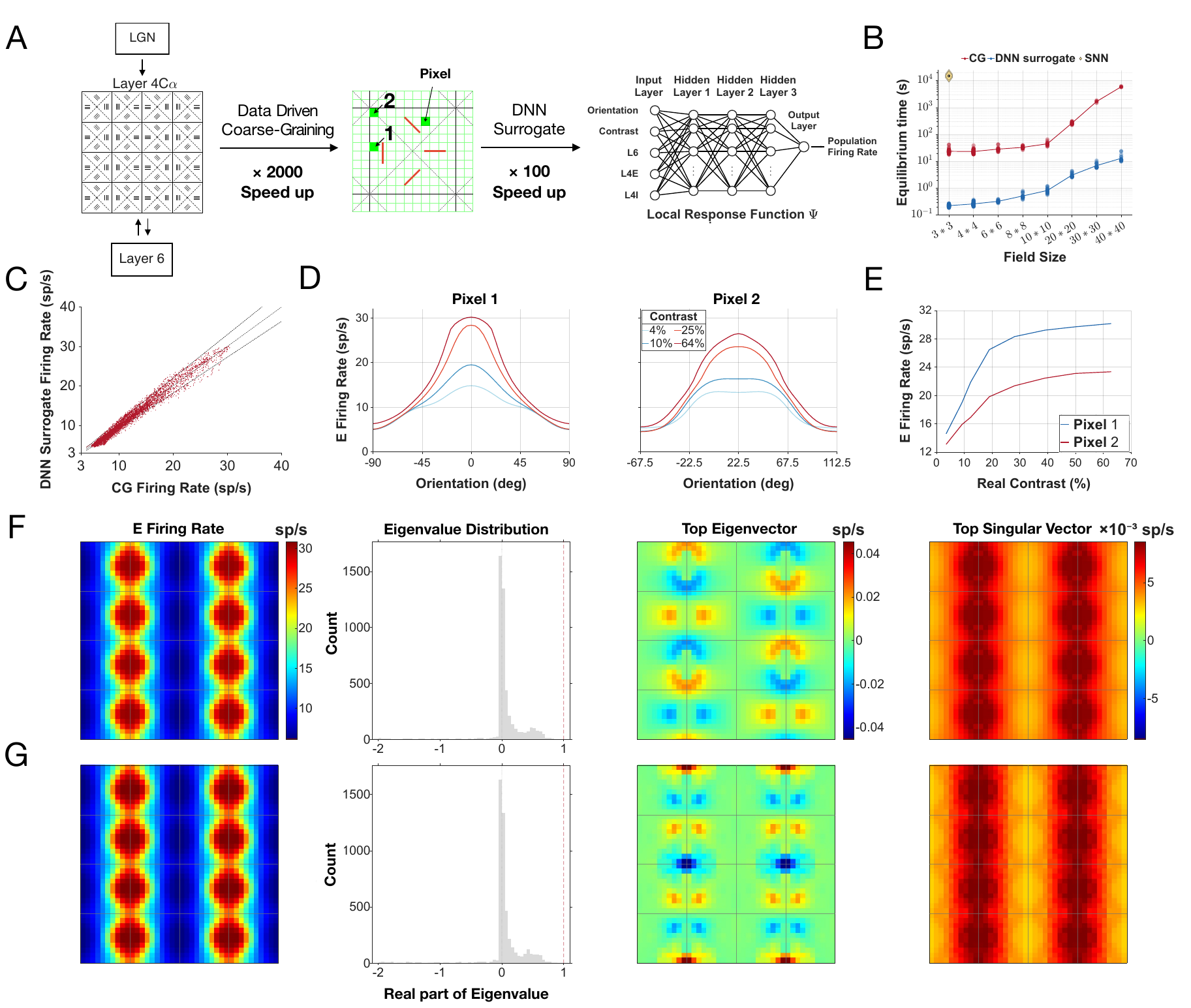}
\caption{\textbf{Deep surrogates accelerate the multiscale V1 model while retaining its visual responses.}
\textbf{A:} Coarse-graining (CG) combines precomputed local neuronal responses across a cortical sheet. Left: The field contains $4\times4$ hypercolumns, and bars indicate assigned orientations; Middle: each HC is divided into $10\times10$ pixels; Right: Four-layer DNNs approximate the local responses of each pixel with inputs from LGN (containing orientation and contrast information), recurrent L4, and L6 feedback.
\textbf{B:} Steady-response computation times, excluding precomputation and training (mean $\pm$ SEM; $n=20$, except the two largest fields: $n=5$). The upper left yellow diamonds show the separate 250-s simulation benchmark for the $3\times3$-HC spiking neural network model.
\textbf{C:} Paired excitatory (E) rates of pixels across all tested visual stimuli; lines indicate equality and $\pm10\%$ error.
\textbf{D,E:} Orientation tuning at contrasts of $4\%$, $10\%$, $25\%$, and $64\%$ (D), and contrast responses (E), for pixels 1 and 2 in A.
\textbf{F,G:} DNN (F) and the parent CG (G) comparisons at $0^\circ$, full contrast drifting grating stimuli. From left to right: E-rate maps, distributions of the real parts of eigenvalues, the slowest right eigenmode, and the leading right singular pattern. The response derivative maps small activity changes to recurrent responses; its eigenmodes describe decay patterns, whereas its leading right singular vector recruits the largest recurrent response.}
\label{fig:1}
\end{figure}

\vspace{0.1in}
\noindent\textbf{Orientation selectivity.} Following the drifting-grating measurements used to characterize macaque V1 tuning\citep{Ringach2002}, we assess OS by recording the excitatory rate of each model pixel as the grating angle is varied. The two pixels marked in Fig.~\ref{fig:1}A prefer approximately $0^\circ$ and $22.5^\circ$, respectively. Pixel 1 lies within an assigned OD, while pixel 2 lies at the boundary between two ODs. Their tuning curves peak near these angles and decrease as the grating is rotated away (Fig.~\ref{fig:1}D). Collectively, the whole L4 circuit exhibit high firing rates at ODs with more preferred angles and low rates at the orthogonal ODs (the first panels of Fig.~\ref{fig:1}F, G). LGN input causes this difference: aligned and orthogonal gratings can deliver different spike patterns to V1 cells even when the average incoming rates are similar. The local response functions preserve this distinction, and cortical interactions combine the selective drive across neighboring populations\citep{Chariker2016,Xiao2024}. An alternative arrangement with six ODs also produces similar, orientation selective responses (Supplementary Fig.~S1).

\vspace{0.1in}
\noindent\textbf{Contrast response.} Contrast measures the light--dark difference in the stimulus. Macaque magnocellular LGN neurons show high sensitivity at low contrast and response saturation at high contrast\citep{DerringtonLennie1984}; low contrast thresholds are also observed in recipient layer 4C$\alpha$\citep{BlasdelFitzpatrick1984}. In the model, increasing contrast changes the feedforward drive and, through recurrent interactions, the cortical inputs received by each population\citep{Chariker2020}. Figure~\ref{fig:1}E shows this steep rise and subsequent saturation at the two example pixels. Increasing contrast also raises their orientation-tuning curves while preserving preferred angle (Fig.~\ref{fig:1}D). 

\subsubsection{Speeding up by coarse-graining and deep surrogates}
\noindent\textbf{Scalability.} The gain in efficiency first comes from reusing the shared potential local responses in the CG model. Local architecture is similar across the field, and each population is driven by a small number of input sources. The previous work therefore chose to compute all potential local responses in advance, then tabulate the response over the relevant range of inputs and reuse this library whenever a local rate is needed \citep{Xiao2024,Xiao2026}. Updating the cortical field requires lookup and interpolation instead of repeating the spiking calculations at every pixel. The same library can also serve different field sizes: a larger cortical sheet requires more interacting pixels, while the local response calculation is unchanged. This separation is the basis of the scalability of the multiscale framework.

Xiao et al.\ 2024 reported a speedup of more than $10^3$ for steady-response evaluation\citep{Xiao2024}.
The library avoids repeated spiking calculations, but its local response functions are still evaluated many times during the tabulation in the CG model. In this paper, we replace the lookup and interpolation step with a DNN approximation.

\vspace{0.1in}
\noindent\textbf{Deep surrogates lead to further speedups and perturbation analysis.}  Three independently parameterized deep neural networks predict the S, C, and I rates. Each receives five inputs: an orientation coordinate defined relative to the local LGN alignment, contrast, L6 drive, recurrent excitation, and recurrent inhibition. Training targets are the local responses in mean firing rates. Once trained, the same functions are used at every pixel.

A fitted function supplies response values between the sampled inputs. This removes the need to populate a dense interpolation grid along every input axis when the local response is regular. Approximation results for particular neural-network classes show how smoothness can improve approximation rates, with further advantages for functions with suitable compositional structure\citep{Yarotsky2017,Poggio2017}. Here the learned functions give an additional median speedup of approximately 100-fold per recurrent update (Supplementary Fig.~S2). Complete equilibrium calculations are also faster across the tested field sizes (Fig.~\ref{fig:1}B). Combined with the earlier $10^3$ CG acceleration, this gives an estimated evaluation advantage of order $10^5$ over spiking simulation after library construction and training (Methods).

The visual responses are retained with this replacement. Across the tested orientation--contrast conditions, paired excitatory rates from the surrogate and parent CG models have a correlation of 0.99 and a mean absolute difference of 0.773 spikes/s; 74.6\% lie within a factor of 1.1 of one another (Fig.~\ref{fig:1}C). The orientation and contrast responses in Fig.~\ref{fig:1}D,E illustrate the functions of the resulting cortical model.

Smooth activation functions provide input derivatives as well as rates. To see their role, consider a small increase in activity at one pixel. The connection kernels linearly convolve the input changes received from other pixels, while the local response slopes specify how much the pooled input rates change the out firing rates.
Combining these steps gives the derivative of the full cortical response. The derivative contributions through recurrent excitation, inhibition, and L6 feedback remain distinguishable. The DNN therefore makes both repeated model evaluation and pathway-specific analysis of perturbations practical.

\subsubsection{Population-rate dynamics and response patterns}

Let $\boldsymbol f=((\boldsymbol f^S)^{\mathsf T},(\boldsymbol f^C)^{\mathsf T},(\boldsymbol f^I)^{\mathsf T})^{\mathsf T}$ collect the local population firing rates across the field, where $\boldsymbol f^Q=(f_p^Q)_{p\in\mathcal P}$ and $\mathcal P$ is the set of all pixels.
Let $\Phi(\boldsymbol{f})$ denote the responses to the inputs generated by this state, assembled from the local response functions $\Psi$ and spatial connectivity kernels $C$ (Methods, Eq.~\ref{Phi}). We introduce a rate equation in which each population adjusts towards its current local response:
\begin{equation}
 \tau\frac{\mathrm d\boldsymbol{f}}{\mathrm dt}=\Phi(\boldsymbol{f})-\boldsymbol{f},
 \qquad \boldsymbol{f}^*=\Phi(\boldsymbol{f}^*).
 \label{eq:model}
\end{equation}
The rate equation supplies a temporal evolution for the population means. Fast spike fluctuations and gamma-phase dynamics are averaged out in computing the local equilibria and are neglected as time-resolved dynamics here. The equation follows coarser changes in firing rates and visual representation. We note that, in our earlier work\citep{Xiao2024,Xiao2026}, we used damped self-consistency update rules to find the steady response, i.e.,
\begin{equation}
\boldsymbol{f}^{(k+1)}=h\Phi(\boldsymbol{f}^{(k)}) + (1-h)\boldsymbol{f}^{(k)}.
 \label{eq:pre-model}
\end{equation}
Equation~\ref{eq:pre-model} has the form of a forward-Euler step for Eq.~\ref{eq:model}, with step size $\Delta t=h\tau$. In the equilibrium solver, however, $h$ is chosen for numerical convergence and the iteration count has no assigned biological duration.

We set the common time constant to $\tau=10.34$ ms by aligning the peak time of maximized transient gain with V1 response timing. For each observation interval, we find the small perturbation that produces the largest response relative to its initial size. The interval of maximum gain is then matched to the empirical timing used for calibration (Methods). The same $\tau$ is held fixed across all pathway and parameter comparisons.

\vspace{0.1in}
\noindent\textbf{Local perturbation analysis.} Near $\boldsymbol{f}^*$, the Jacobian matrix $J=D_{\boldsymbol{f}}\Phi(\boldsymbol{f}^*)$ describes the recurrent response to a small activity change. We focus on the \textit{singular modes} and \textit{eigenmodes} obtained from the Jacobian.

The leading right singular vector is the perturbation direction that recruits the largest recurrent response; the leading left singular vector is its output pattern, and the singular value is their size ratio (Methods). These patterns describe the recurrent component of the transient response of L4 to perturbations. Its time course also includes population relaxation and repeated feedback. An example trajectory illustrates how a perturbation of $\boldsymbol{f}^*$ can be amplified before returning to equilibrium
(Supplementary Fig.~S9).

Eigenvectors identify modes that the recurrent response returns as scalar multiples of themselves. For a mode $v$ with eigenvalue $\lambda$, the linearized rate equation gives
\begin{equation}
 Jv=\lambda v,\qquad
 \delta \boldsymbol{f}(t)=e^{(\lambda-1)t/\tau}v.
 \label{eq:linear}
\end{equation}
For real eigenvalues this describes a fixed spatial pattern with changing amplitude; complex eigenvalues also describe oscillation. The amplitude decays when the real part of $\lambda$ is below one and grows when it is above one. Therefore, eigenvalues with real parts just below one describe slowly decaying or persistent patterns, whereas a real part above one implies instability of $\boldsymbol f^*$. The vertical line at one in the spectral plots is thus the stability boundary of the rate model.

The parent CG model and DNN surrogate show similar spectral organization: many eigenvalues lie near zero, while a smaller group extends towards the stability boundary (Fig.~\ref{fig:1}F,G). Their leading eigenmodes occupy similar regions of the cortical field. Closely spaced eigenvalues can make individual eigenvectors sensitive to small derivative changes\citep{Greenbaum2020}; we therefore also examine groups of leading modes and their spatial organization. The broad spectral organization persists across pixel resolutions (Supplementary Fig.~S4). We now use the pathway decomposition to determine which inputs sustain the slow patterns and which stabilize them.

\subsection{Layer-6 feedback supports slow recovery, and inhibition stabilizes the response}
\label{sec:stability}

\begin{figure}[p]
\centering\includegraphics[width=0.6\textwidth]{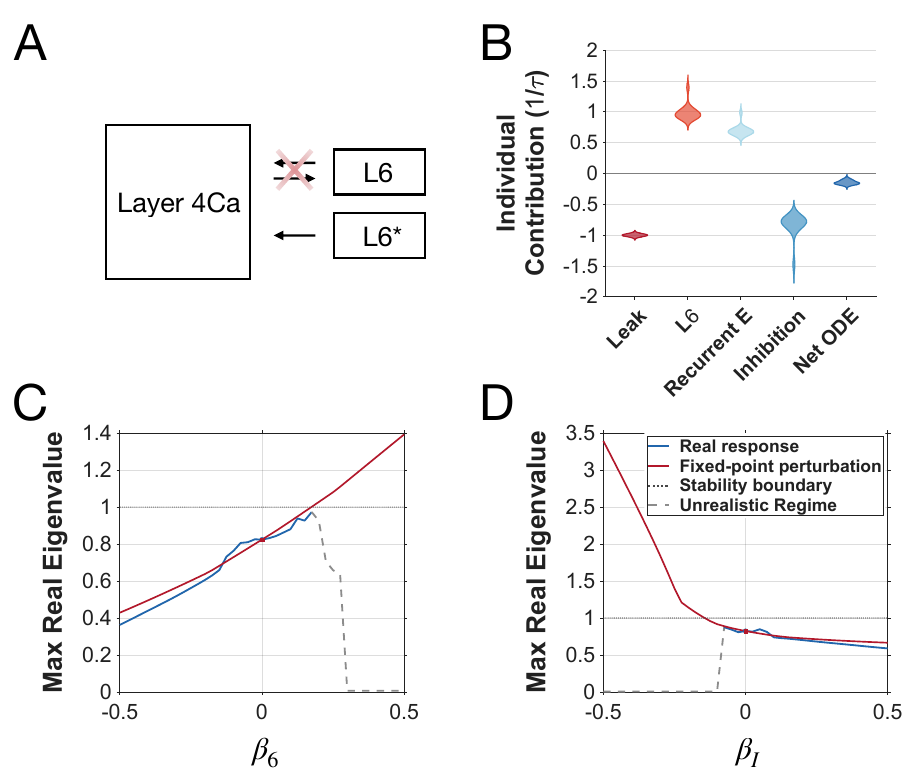}
\caption{\textbf{L6 feedback sustains slow modes, while inhibition promotes their decay.}
\textbf{A:} A fixed-point-preserving (FPP) intervention mixes activity-dependent input with its equilibrium value, changing feedback while preserving the fixed point. L6* denotes the FPP input mixture.
\textbf{B:} Contributions to the slowest mode's growth or decay across all orientation--contrast combinations of visual stimuli. Positive values sustain activity; negative values promote decay. Leak denotes population relaxation; $\tau$ is the population time constant.
\textbf{C,D:} The largest response-derivative eigenvalues under changes of pathway strength in L6 ($\beta_6$) or L4 inhibition ($\beta_I$). Blue curves correspond to real response to pathway strength changes (which moves the fixed point), while red curves use matched FPP interventions. Values above one imply instability. Grey dashed segments mark unrealistically high-rate calculations excluded from the comparison.}
\label{fig:2}
\end{figure}

We first ask how individual pathways determine the stability and recovery of a given visual response. Changing a pathway's strength usually changes the steady firing-rate map as well, so we initially hold that map fixed while altering the pathway's response to a disturbance. For a selected pathway, we replace a chosen fraction of its activity-dependent input by the constant input it supplies at that response (Fig.~\ref{fig:2}A). The original response remains an exact equilibrium, while the pathway's reaction to an activity change becomes weaker. We call this a \textit{fixed-point-preserving (FPP) intervention}.

For example, consider a pixel receiving L6 feedback. At the reference response, freezing this pathway supplies exactly the same input as before. If the pixel's activity is then displaced, the frozen L6 input stays constant, whereas the original feedback would change with L4 activity. The comparison asks how feedback generated by the disturbance affects its recovery. The same operation can be applied to recurrent excitation and inhibition, or restricted to any selected pathways and pixels. 
{We define the pathway change index $\beta$: $\beta=-1$ means the selected pathway is completely frozen (so the strength decrease for one time), while $\beta=0$ means no intervention applied and the system stays intact. Similarly, all other values of $\beta$ indicate linear interpolations/extrapolations of the input from the corresponding pathway of these two scenarios. For example, $\beta=1$ hence means the pathway strength is doubled, though a fixed, negative surrogate input from the same pathway is provided to preserve the fixed point. In general, $\beta<0$ means decrease the pathway strength and vice versa.}

\begin{figure}[p]
\centering\includegraphics[width=\textwidth]{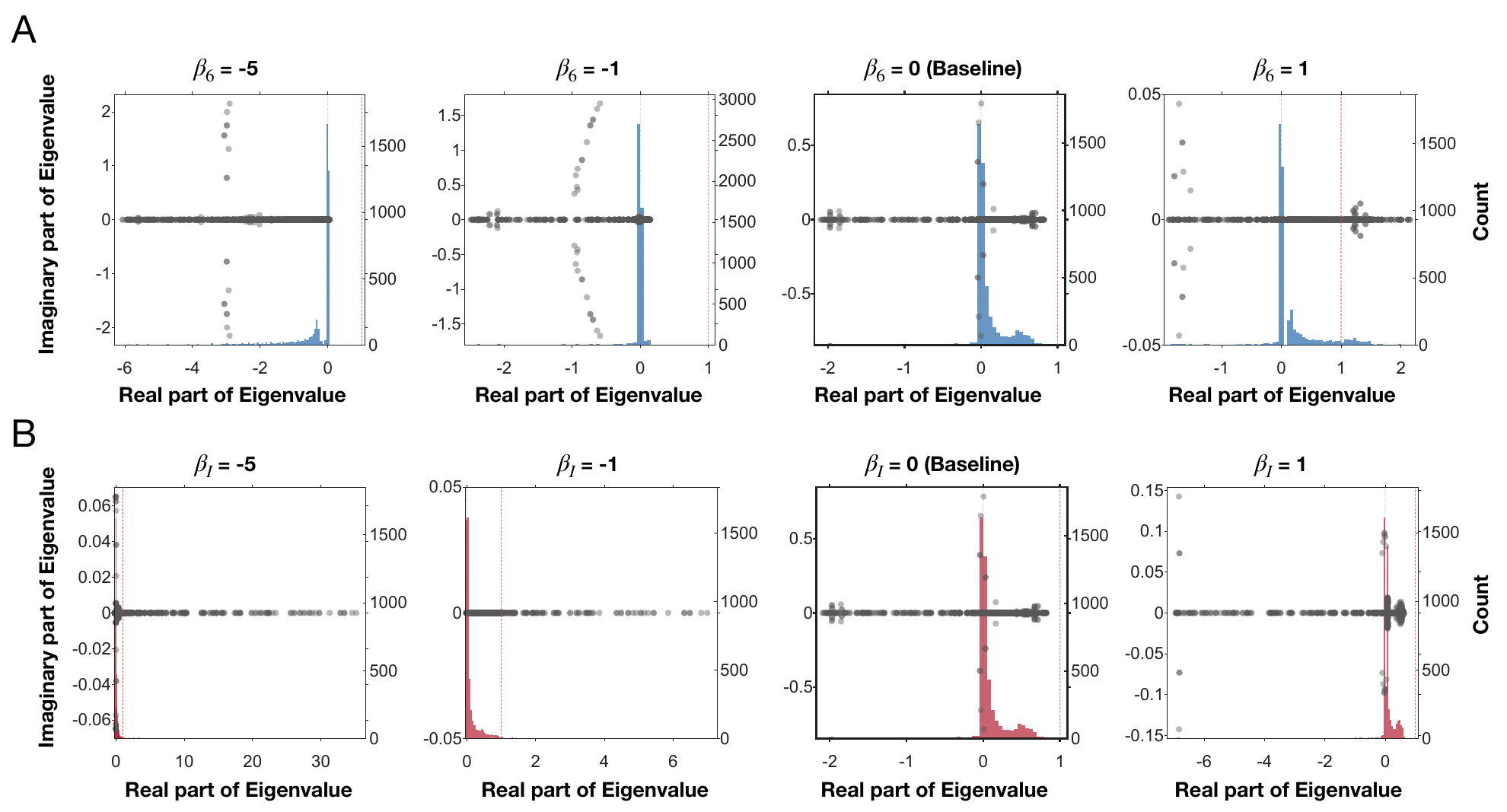}
\caption{\textbf{L6 feedback and inhibition shape different parts of the spectrum.}
Response-derivative spectra under fixed-point-preserving changes to L6 feedback (A) or inhibitory input to all populations (B). Columns use $\beta=-5,-1,0,1$, giving activity-dependent multipliers $-4,0,1,2$: sign-reversed feedback, frozen, baseline, and doubled. The fixed point is preserved throughout. Grey points show eigenvalues; blue and red histograms count their real parts. The red dashed line at $\mathrm{Re}=1$ marks the rate-model stability boundary. Stronger L6 feedback or frozen inhibition permits unstable modes. Axis ranges differ.}
\label{fig:3}
\end{figure}

We calculate each pathway's signed contribution to the growth rate of the full model's slowest mode (Methods). Positive contributions help sustain the pattern; negative contributions promote its decay. At each tested contrast and orientation, L6 feedback and recurrent excitation make positive contributions, while inhibition and population relaxation make negative contributions (Fig.~\ref{fig:2}B). Their sum is small and negative, so the pattern decays slowly. Freezing L6 feedback moves the leading eigenvalue away from the stability boundary ($\beta_6=-1$, Fig.~\ref{fig:3}B). Freezing inhibitory input has the opposite effect: originally negative growth rates become positive, corresponding to eigenvalues of $J$ beyond that boundary ($\beta_I=-1$, Fig.~\ref{fig:3}B). These relationships hold across all tested orientation--contrast conditions (Supplementary Fig.~S3). In general, the response derivative is the sum of contributions carried by these pathways.

This is the mechanism of inhibitory stabilization. Feedback inhibition stabilizes a strongly recurrent excitatory circuit. Experimental evidence consistent with this regime comes from intracellular surround-suppression measurements in cat V1\citep{Ozeki2009}; related models explain visual responses through recurrent excitation and inhibition\citep{Rubin2015}. Here freezing inhibitory output exposes an unstable excitatory subsystem that includes the effective L6 recurrence (Methods). L6 feedback provides excitatory support to the slow modes, consistent with the parent model's prediction of net excitatory L6 input to L4 in macaque\citep{Chariker2020}. In mouse V1, optogenetic activation of L6 corticothalamic neurons can instead suppress activity in other cortical layers by recruiting inhibition\citep{Olsen2012,Bortone2014}.

We also compare these controlled interventions with actual changes in pathway strength, for which the equilibrium is recomputed rather than freezing any input (Fig.~\ref{fig:2}C,D). Both computations show that stronger L6 feedback raises the leading eigenvalue and stronger inhibition lowers it. The fixed-point-preserving intervention retains the reference equilibrium mathematically even after it loses stability, but that equilibrium then ceases to be locally attracting. At the grey dashed parameter settings, the pathway-strength calculations enter a regime of unrealistically high firing rates, marked by the grey dashed segments in Fig.~\ref{fig:2}C,D.

L6 and L4 inhibition affect different parts of the spectrum. Increasing L6 feedback moves the leading modes (with real parts close to one) towards, and eventually across, the stability boundary ($\beta_6 = 1$, Fig.~\ref{fig:3}A). Reversing its weight it moves these eigenvalues to negative while leaving much of the concentration near zero in place ($\beta_6 = -5$, Fig.~\ref{fig:3}A). On the other hand, increasing inhibition extends the negative part of the spectrum ($\beta_I = 1$, Fig.~\ref{fig:3}B), while freezing it permits a broad positive extension ($\beta_I = -1$, Fig.~\ref{fig:3}B). Thus L6 support for slow modes and inhibitory stabilization coexist with a large group of modes receiving little recurrent support.

Aside from the leading modes and the negative spectrum, another persistent feature of the spectrum is the large group of near-zero eigenvalues. We next examine their origin.

\subsection{Spatial pooling and pathway cancellation shape near-zero recurrent responses}

\begin{figure}[p]
\centering\includegraphics[width=0.7\textwidth]{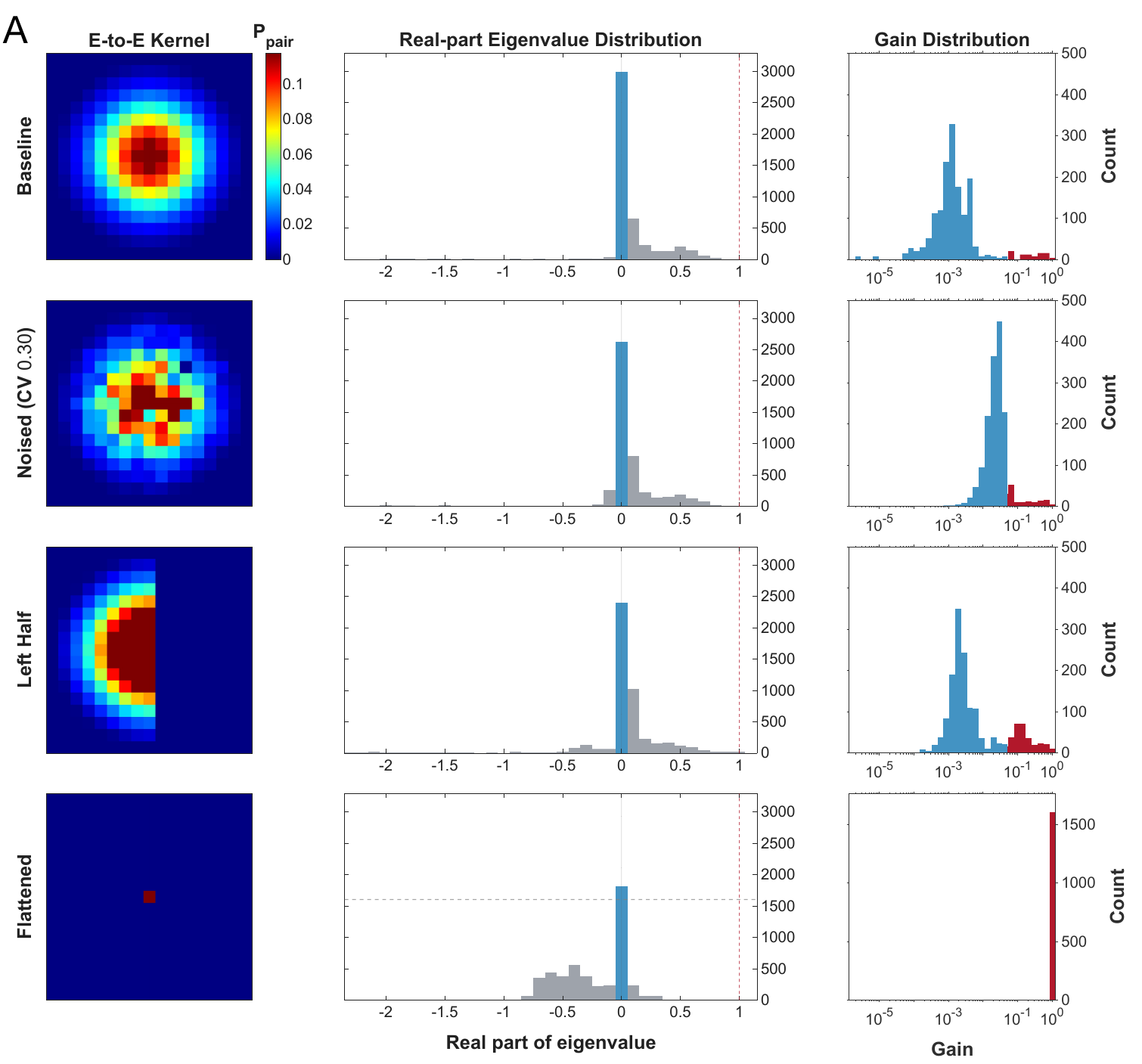}
\caption{\textbf{Spatial pooling attenuates fine-scale activities.}
Rows compare baseline, noisy (coefficient of variation 0.30), left-half, and locally concentrated (``Flattened'') connections. Left: representative excitatory-to-excitatory kernels on a shared color scale. Middle: distributions of all 4,800 response-derivative eigenvalues by real part; blue marks the near-zero bin $[-0.05,0.05)$. Vertical lines mark zero and the stability boundary at one; the horizontal dashed line marks 1,600 eigenvalues. Right: gains for the kernel's 1,600 spatial Fourier modes, normalized to the gain for a uniform activity change. Blue and red indicate gains $\leq0.05$ and $>0.05$. Wide and symmetric kernels average out most spatial modes, giving rise to many near zero eigenvalues, whereas concentrated kernels retain all spatial modes, without increasing near zero eigenvalues.}
\label{fig:4}
\end{figure}

Many eigenvalues cluster near the origin, indicating weak recurrent support for their corresponding eigenmodes (Fig.~\ref{fig:4}). Modes with real parts near zero decay on a time scale close to $\tau$, in contrast to the slowly decaying modes near the stability boundary. We examine two circuit features that can make this happen: spatial pooling and excitation-inhibition balance (or E-I balance) between pathways.

\vspace{0.1in}
\noindent\textbf{Spatial pooling.} The Gaussian-like spatial connectivity kernel makes a target population less sensitive to fine variation across neighboring pixels. Broad, coherent activity changes contribute with the same sign and largely survive pooling. In contrast, rapid spatial alternations contribute with opposite signs and largely cancel. The connection kernel's Fourier transform quantifies this difference by assigning a gain to each spatial frequency. The extended baseline kernel has low gains for most high-frequency spatial modes, indicating that many possible input patterns are strongly attenuated (Fig.~\ref{fig:4}, top row; Methods). Related ideas appear in classical orientation-network models and low-rank network theory, where organized connectivity supports selected population patterns\citep{BenYishai1995,Mastrogiuseppe2018}.

To test the role of pooling, we alter the spatial kernels while preserving their total incoming strength, then recompute the equilibrium and its derivative. Adding noise to the baseline kernel or retaining only its left half preserves a prominent near-zero eigenvalue concentration and low Fourier gains for many spatial modes (Fig.~\ref{fig:4}, the second and third rows). In contrast, concentrating the connections at the target pixel removes spatial pooling. In this control, labeled ``Flattened,'' the kernel transmits all spatial frequencies equally and the near-zero eigenvalue concentration is reduced (Fig.~\ref{fig:4}, bottom row; See also Fig.~S5 for additional kernel comparisons). Removing L4 spatial pooling also reduces the number of weak-gain singular directions and increases the effective rank of the response-map Jacobian, supporting spatial pooling as a source
of weak recurrent responses (Supplementary Fig.~S10).

\vspace{0.1in}
\noindent\textbf{E-I balance.} Pathway cancellation provides a second source of weak recurrent responses. For a selected inhibitory activity change, we choose an accompanying excitatory change that cancels the recurrent response in the excitatory populations. We then examine the remaining response in the inhibitory populations, comparing the direct inhibitory contribution with the contribution mediated by the accompanying excitatory change (Methods, ``Spatial kernels and excitatory--inhibitory elimination''). Many tested patterns show strong relative cancellation, including patterns with broad spatial variation (Supplementary Fig.~S7). Thus opposing pathway contributions can reduce recurrent responses even when spatial pooling alone would transmit the pattern.

In addition, combining simple and complex cells into one excitatory population removes a further group of near-zero eigenvalues while retaining much of the remaining spectrum (Supplementary Fig.~S6). Together, pooling and pathway cancellation help explain why the recurrent circuit responds weakly to many population changes. We note that pathway effects cancel for selected activity perturbations at fixed circuit parameters, whereas the later compensation analysis varies pathway strengths together to preserve steady responses approximately.

\subsection{Slow patterns and strong recurrent responses have different spatial profiles}
\begin{figure}[p]
\centering\includegraphics[width=\textwidth]{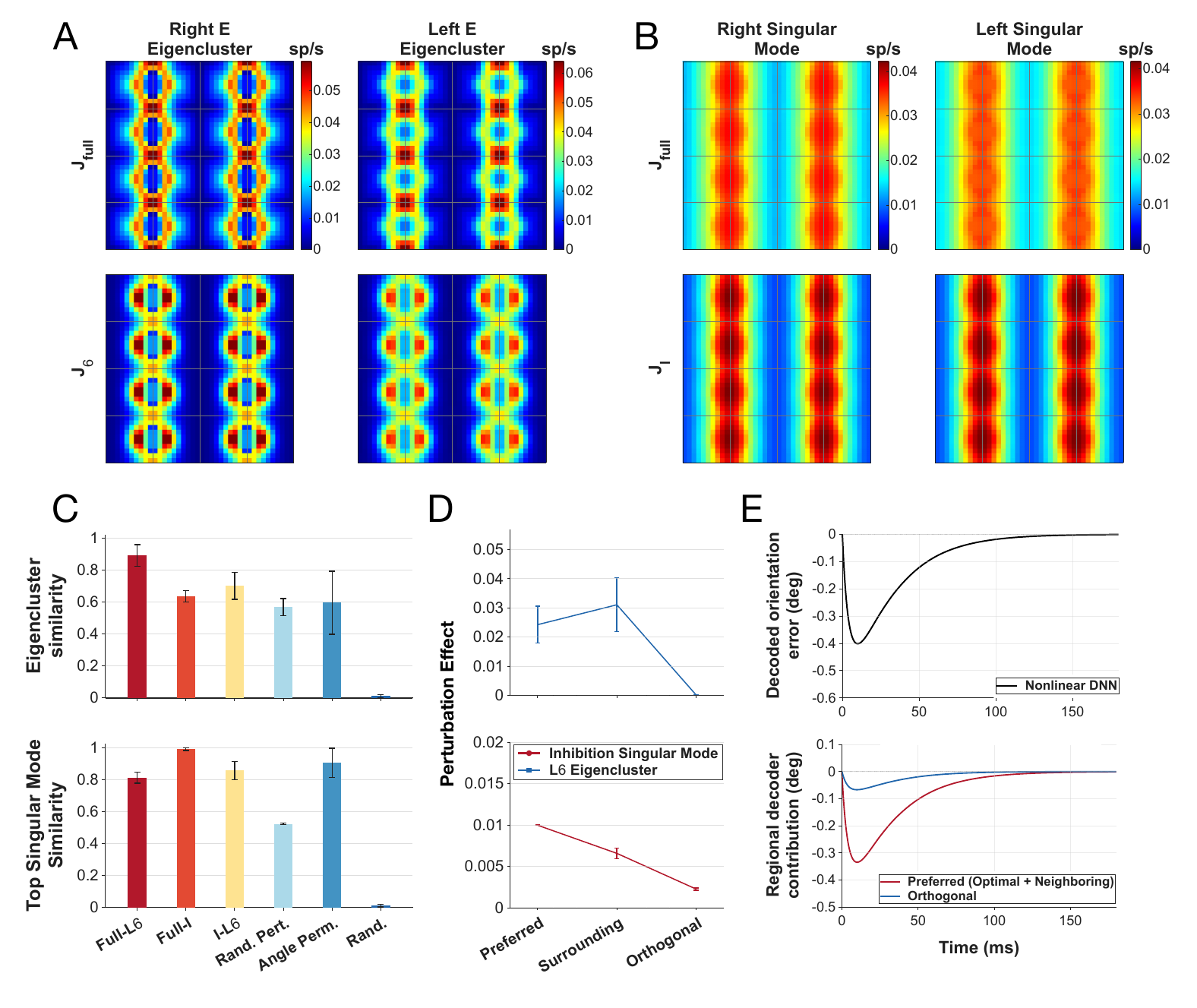}
\caption{\textbf{Pathway-specific spatial patterns and recovery of the represented orientation.}
The Jacobians $J$, $J_6$, and $J_I$ denote the full response derivative and its L6 and inhibitory contributions.
\textbf{A:} Envelopes of the leading eigenmodes (eigenclusters) of $J$ and $J_6$ (showing the spatial map on E firing rates). The leading modes of $J$ describe slow recovery in the full circuit. Right envelopes show spatial support of these patterns; left envelopes summarize sensitivity to perturbations.
\textbf{B:} Inhibitory components of the perturbation producing the largest recurrent response at fixed size and its output (right/left singular vectors of $J$) in the full circuit and inhibitory pathway. Maps are normalized separately.
\textbf{C:} Cosine similarities between spatial patterns (mean $\pm$ SD; across all visual stimuli), with added-noise, orientation-permuted, and completely random controls (Methods). Upper panel compares full circuit/L6; lower panel compares full circuit/inhibition. 
\textbf{D:} Profiles of the L6 eigencluster (top) and inhibition-associated singular pattern (bottom) across preferred, neighboring, and orthogonal domains (mean $\pm$ SEM, four orientations); the two families have separate normalizations.
\textbf{E:} Following a perturbation, the local orientation readout briefly shifts despite stimulated by a fixed grating (nonlinear DNN trajectory). The readout removes first-order gain and contrast changes. Lower curves partition the readout into preferred-plus-neighboring and orthogonal-domain contributions.}
\label{fig:5}
\end{figure}

We next examine the spatial organization of slowly decaying eigenmodes and the singular patterns that recruit the strongest recurrent responses. To account for the contributions from different pathways, we split the full Jacobian $J=J_6+J_E+J_I$, where each term is the derivative carried through L6 feedback, recurrent excitation, or recurrent inhibition, evaluated at the same reference response (Methods, Eq.~\ref{eq:pathway_derivatives}). We report that, across the tested visual stimuli, the full model's slow patterns closely resemble the leading patterns of $J_6$, whereas its strongest recurrent output patterns more closely resemble those of $J_I$ (Fig.~\ref{fig:5}A--C).

Several leading modes of the full model have similar decay rates. We group these modes into an eigencluster and summarize their spatial magnitudes by an envelope, applying the same construction to the leading modes of each pathway contribution (Methods). The right envelope shows where the leading patterns are expressed; the left envelope summarizes the spatial support of their sensitivity to a disturbance. The full-system and L6 envelopes occupy similar regions around active ODs and show almost zero activity at the orthogonal ODs (Fig.~\ref{fig:5}A). Their cosine similarity is higher than that of the full--I and I--L6 pairs. Note that the full-I and I-L6 pairs are not significantly different from the 0.6 similarity when permuting stimulus orientations (Fig.~\ref{fig:5}C, upper panel). This close correspondence agrees with the pathway interventions in Section 2: effective L6 feedback sustains the slow modes and moves them towards instability when strengthened.

The leading singular patterns show a different correspondence. Figure~\ref{fig:5}B displays the inhibitory components of the perturbation patterns that recruit the largest recurrent responses for a fixed initial size, together with their output patterns. The output pattern of the full system closely resembles that of L4 inhibition, as also reflected in the cross-stimulus similarity comparison (Fig.~\ref{fig:5}C, lower panel). Inhibitory feedback can therefore generate a strong transient response while stabilizing the circuit longer term. 

The spatial profiles also distinguish the two pathway-associated patterns (Fig.~\ref{fig:5}D). The L6 eigencluster envelope is concentrated in neighboring rather than the most preferred ODs, with little amplitude at the orthogonal domain. The inhibition-associated singular pattern extends more broadly. 

The pathway-freezing controls show that layer 6 and inhibition also alter finite-time gain and subsequent decay (Supplementary Fig.~S8; Methods). The equilibrium patterns therefore identify pathway contributions that can be examined in the evolving response.

\vspace{0.1in}
\noindent\textbf{Transient decoding error after perturbation.} Perturbations of the firing-rate map also produce transient errors in the orientation read out from the population, even while the stimulus remains fixed (Fig.~\ref{fig:5}E). We measure small orientation shifts around the reference stimulus with a readout that removes first-order gain and contrast changes (Methods). Both the slowly decaying patterns and the strong recurrent-response patterns are concentrated in preferred and neighboring ODs, with weaker expression in orthogonal ODs. Consistently, most of the decoding deviation during recovery comes from the preferred and neighboring domains, while the orthogonal-domain contribution is small. The spatial profiles and readout thus suggest that the tested disturbances mainly distort the representation near the presented orientation, rather than strongly recruit an orthogonal-orientation pattern.

\subsection{Pathway compensation preserves broad tuning while changing recovery}
In the background regime without visual input, our previous model study found that changes as small as 1\% in some recurrent synaptic weights could move firing rates outside the predefined biologically plausible range. Coordinated parameter changes nevertheless defined a thin viable region, indicating that sensitivity to individual parameters can coexist with flexibility under compensation\citep{xiao2021data}. Here we ask whether changes in pathway strengths can offset one another in the steady response, producing different circuits with similar firing-rate maps under visual stimulation.

We measure approximate preservation by the weighted root-mean-square change across the aggregate excitatory and inhibitory firing-rate maps, using the HC norm to measure the weighted firing rate difference across one hypercolumn\citep{Xiao2024} (Methods, Eq.~\ref{eq:HC}). A minimum-error direction identifies the best sampled combinations in each parameter plane. The preferred compensation direction also depends on the observable we choose.

\begin{figure}[p]
\centering\includegraphics[width=\textwidth]{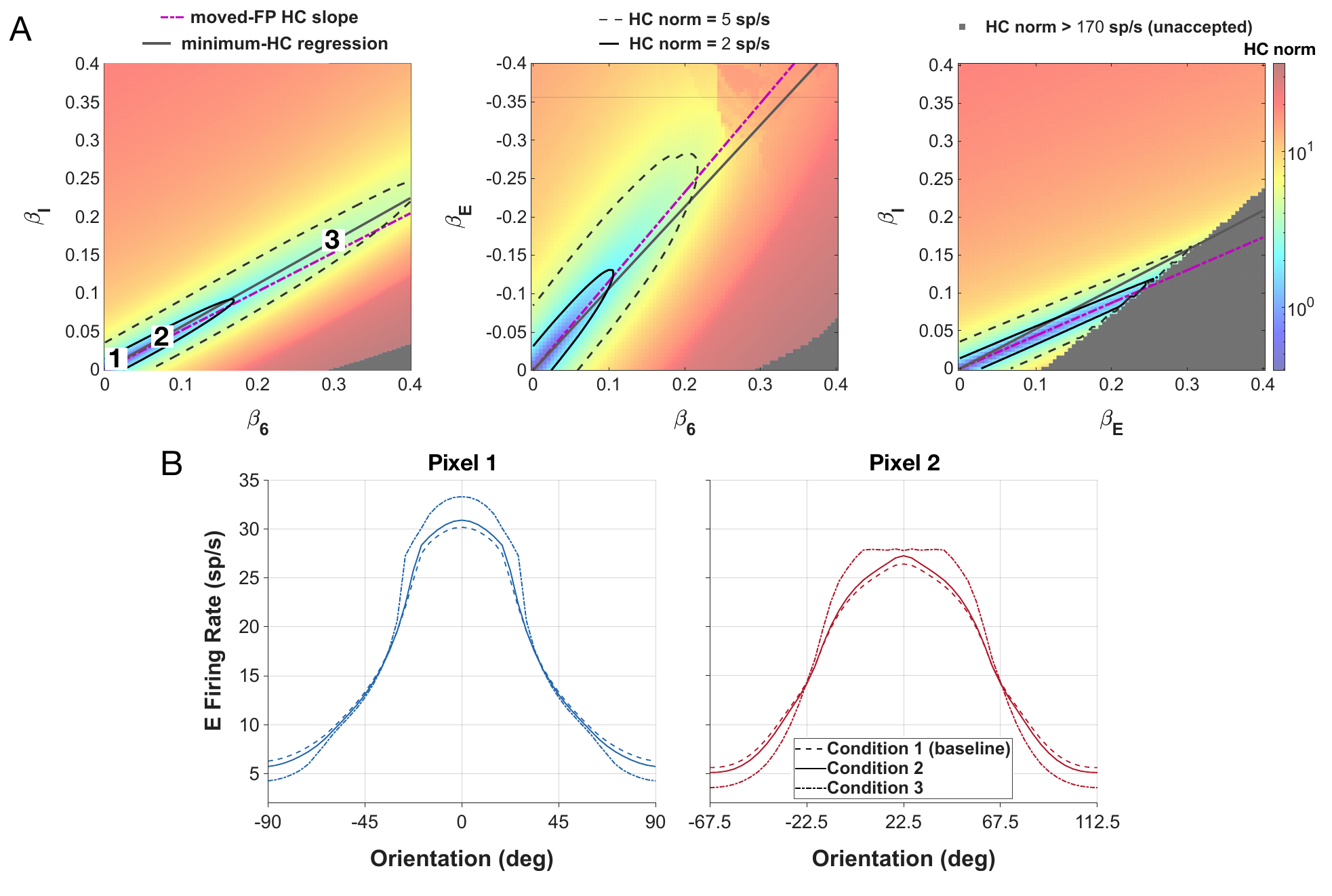}
\caption{\textbf{Coordinated pathway changes preserve broad orientation tuning.}
\textbf{A:} Pairwise changes in L6 feedback, recurrent excitation, and inhibition. Parameters $\beta_6$, $\beta_E$, and $\beta_I$ are fractional strength changes from baseline. Colors show response error in the HC norm, a weighted root-mean-square change in excitatory and inhibitory rates (logarithmic scale). Magenta lines predict local compensation; grey lines fit numerical minimum-error paths. Black solid/dashed contours mark 2/5 spikes/s; grey regions exclude errors above 170 spikes/s.
\textbf{B:} Orientation tuning at pixels 1 and 2 (Fig.~\ref{fig:1}A) for the three marked conditions along the L6--inhibition compensation line in A: baseline (dashed), Condition 2 (solid), and Condition 3 (dash-dotted). Broad preferences persist despite changes in response amplitude.}
\label{fig:6}
\end{figure}

Figure~\ref{fig:6}A compares three pairs of pathway changes. Increasing L6 drive can be offset by increasing recurrent inhibition onto all populations or by decreasing recurrent excitation. This is because, as discussed in Section~\ref{sec:stability}, the effective L6 pathway in this model is net excitatory to L4 while exciting both L4E and I cells. Similarly, increasing recurrent excitation can be offset by increasing inhibition. Each plane contains a narrow band of pathway combinations with small differences between firing-rate maps: strengths can vary by approximately 15--40\% before the HC difference reaches 5 spikes/s. The direction predicted from the derivative of the steady response agrees approximately with the numerical minimum-error path near baseline (Methods).

These bands arise from coordinated changes whose effects overlap across the circuit. They provide a spatially resolved example of the parameter multiplicity found in other neuronal models\citep{Prinz2004}. Here, compensation gives us alternative circuits with similar steady responses in which to compare visual tuning and recovery dynamics. Nevertheless, we note that the goodness of compensation depends on which pathways are varied: Supplementary Fig.~S11 shows an example where two parameters do not effectively compensate each other. This is because the L6 pathway excites both L4E and L4I cells, whereas changing the I-to-E pathway alters inhibition onto L4E but not L4I. The resulting effects on the two populations therefore compensate less effectively.


We follow the fitted L6--inhibition compensation line, where a 10\% increase in L6 drive is approximately paired with a 5.6\% increase in inhibition. The three marked conditions comprise baseline and two compensated circuits (Fig.~\ref{fig:6}A, left). Their tuning curves at two representative pixels retain broad orientation preferences and similar shapes, with modest changes in amplitude (Fig.~\ref{fig:6}B). Across these three conditions, the equilibrium E-rate pattern, leading singular-input pattern, and the spatial envelope of the leading eigencluster remain similar (Supplementary Fig.~S12).

\begin{figure}[p]
\centering\includegraphics[width=\textwidth]{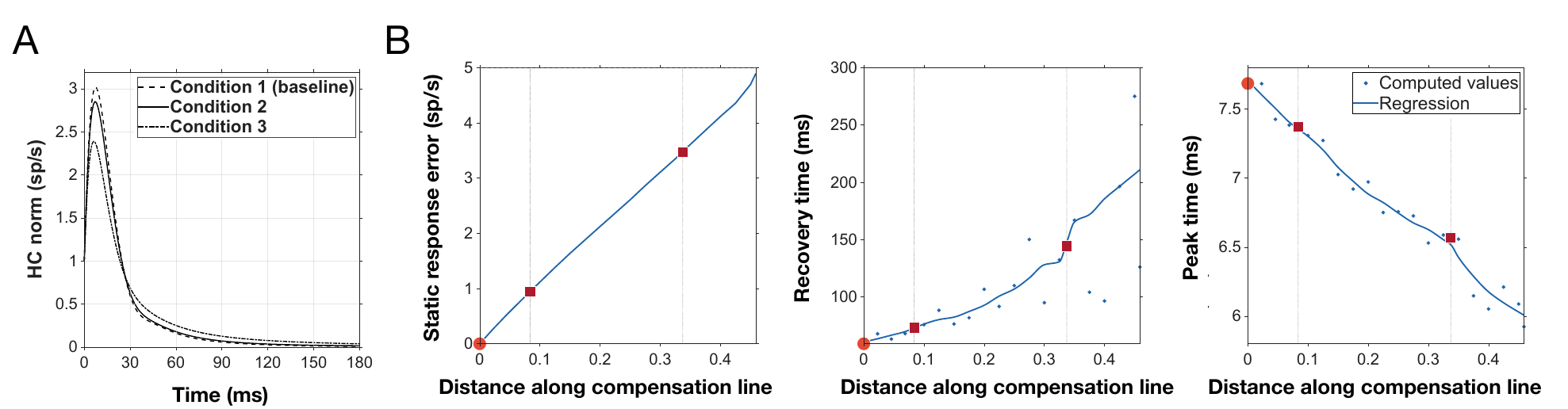}
\caption{\textbf{Similar tunings with different transient responses and recovery times.}
\textbf{A:} Nonlinear responses to the same initial displacement, of HC size 1 spike/s, from each circuit's own equilibrium (initialization protocol: Methods, ``Response error and perturbation size''). Conditions and line styles match Fig.~\ref{fig:6}. All responses amplify temporarily before decaying, with different peak amplitudes.
\textbf{B:} Measures along the L6--inhibition compensation line, $\beta_I\simeq0.562\beta_6$; horizontal axes give distance from baseline in parameter space. Left: equilibrium response error in the HC norm. Middle: the slowest modal recovery time, determined by the leading eigenvalue. Right: time of maximum response to the fixed perturbation. The red markers indicate conditions 1-3 in Fig. 6A. Modal recovery time varies in a wide range, while the transient peak occurs modestly earlier. Dots show computed values, and solid curves show locally weighted linear regression. Definitions and temporal sampling are given in Methods.}
\label{fig:7}
\end{figure}

These compensated circuits nevertheless differ in their recovery dynamics. We apply the same initial displacement, with HC norm 1 spike/s, relative to each condition's own equilibrium. All three nonlinear responses initially amplify the disturbance and then decay, but their peak amplitudes and subsequent trajectories differ (Fig.~\ref{fig:7}A). Each trajectory is measured from its own equilibrium, so these differences describe the response to the disturbance rather than a difference between steady firing-rate maps.

With the common model time scale held fixed, the slowest modal recovery time along the compensation line varies from roughly 60 to 280 ms, whereas the peak time of the response to a fixed perturbation decreases more modestly, from approximately 7.5 to 6 ms (Fig.~\ref{fig:7}B). The former is determined by the leading eigenvalue; the latter depends on the chosen perturbation and the evolving combination of modes (Methods). A slowly decaying mode need not be strongly recruited by every perturbation, so its time constant and the duration of a particular trajectory describe different aspects of recovery.

The controlled pathway analysis explains how these changes can coexist. Pathway changes whose effects on steady firing rates approximately cancel can still alter the feedback evoked by a disturbance. Compensation of the steady response therefore does not imply compensation of recovery dynamics. Joint measurements of steady responses and perturbation recovery could distinguish pathway combinations that remain similar under the steady-response criterion alone.

\section*{Discussion}

Perturbation dynamics provide constraints on cortical circuits that steady visual responses alone leave unresolved. In the present model, effective L6 feedback and recurrent excitation sustain slowly decaying activity patterns, while recurrent inhibition maintains stability. The spatial structure of the strongest recurrent responses closely resembles that of the inhibitory pathway, whereas the slow patterns resemble those supported by L6 feedback. These are complementary pathway contributions to a coupled circuit: amplification and recovery over time depend on their interaction. Spatial pooling and opposing pathway effects also make many activity changes weakly effective as recurrent inputs.

The compensation results show why these dynamical distinctions matter for circuit inference. Coordinated changes in pathway strengths can approximately preserve aggregate E/I firing-rate maps and broad orientation tuning while changing transient amplitudes and slow recovery times. The effects on steady firing rates can cancel without preserving the feedback evoked by a disturbance. This extends the use of multiscale models from testing circuits through their steady responses\citep{Xiao2024,Xiao2026} to comparing how candidate circuits recover. Joint steady-response and perturbation measurements could therefore distinguish pathway combinations that satisfy the same steady-response criterion.

\vspace{0.1in}
\noindent\textbf{Experimental tests of recovery and pathway contributions.} High-density laminar electrophysiology can be used to measure recovery dynamics in macaque V1\citep{Trepka2022}. Comparing responses to weak perturbations under a fixed visual stimulus can reveal transient amplification, slow decay, and shifts in a population orientation readout. Each recording samples a limited subset of L4C$\alpha$ neurons, providing a local view of these dynamics. Conventional calcium imaging offers denser spatial sampling, but its temporal resolution and limited access to deep L4C$\alpha$ make electrophysiology more suitable for the rapid recovery measurements considered here\citep{Tang2018}.

Pathway contributions could be examined by modulating inhibitory drive within L4 and the activity of L6. GABAergic optogenetic manipulation is feasible in macaque V1\citep{De2020} and offers an approach to varying inhibitory recruitment, while targeted modulation of L6 activity could probe the contribution of deep-layer feedback. Comparisons across conditions with similar steady orientation and contrast responses would test whether altered pathway contributions produce different transient amplification and subsequent decay. These measurements would connect the model's predictions of inhibitory stabilization and L6-supported persistence to experimentally accessible changes in circuit activity.

Initial model--experiment comparisons should focus on coarse spatial summaries of population activity, because detailed population rates across the modeled cortical sheet are difficult to sample simultaneously. Spatially pooled responses and their recovery trajectories can already help distinguish candidate circuits with similar steady tuning. Testing the finer spatial organization of slow modes and amplified patterns would require denser recordings distributed across orientation domains, while resolving the fastest predicted transients would require higher effective temporal resolution. Together, these measurements would progressively constrain how different pathway combinations support similar visual functions.

\vspace{0.1in}
\noindent\textbf{Computational scope and limitations.} DNN approximations of local spiking-population responses accelerate recurrent updates by approximately a factor of 100 and provide derivatives for pathway-specific perturbation analysis. Combined with the earlier CG acceleration\citep{Xiao2024}, the estimated evaluation advantage is of order $10^5$ over spiking simulation, based on separate benchmarks and excluding library construction and training. The added population-rate equation describes the evolution of coarse firing-rate statistics; agreement of steady responses does not by itself validate its transient dynamics. The present stability analysis is local to the tested responses and does not exclude other attractors. The recovery times also depend on the common time calibration, which is held fixed across pathway comparisons.

\vspace{0.1in}
\noindent\textbf{Future Directions.} 
A natural extension is to test whether these pathway relationships persist across spatial frequency, temporal frequency, and motion direction as well as orientation and contrast. Experimental observations show that these features are not represented independently: orientation and spatial-frequency discrimination deteriorate near detection threshold\citep{Webster1990}; temporal frequency can alter direction selectivity in ferret V1\citep{Moore2005}; and contrast changes joint spatial- and temporal-frequency tuning in macaque V1 complex cells\citep{Priebe2006}. The local response derivatives and recurrent connections determine how small stimulus changes reshape the population response. Extending the present analysis across stimulus families could reveal which pathway combinations preserve feature distinctions and which amplify disturbances that obscure them.

The same approach can be extended to multiple V1 layers and higher visual areas. Each CG update combines spatial pooling with nonlinear local responses, providing a biologically constrained architecture for comparison with recurrent convolutional networks. Artificial neural network models trained on recordings already predict macaque V1 responses to natural images\citep{Cadena2019}, and a foundation model trained on natural-video responses predicts orientation and direction tuning in mouse visual cortex\citep{Wang2025}. Building and comparing larger mechanistic models will require testing many uncertain parameters and architectures. Fast CG evaluation can reduce the cost of each test, while generative methods can amortize parameter inference across target responses through a shared simulation library\citep{Zhang2026Diffusion,Goncalves2020}. Comparing both steady representations and recovery under matched perturbations would then test whether similar visual functions are supported by similar circuit mechanisms.

\section*{Methods}

\subsection*{Model organization and local response functions}

The model inherits the local-population construction and spatial interactions of the earlier multiscale models\citep{Xiao2024,Xiao2026}. The main cortical field contains $N=1600$ pixels on a periodic $40\times40$ grid. Each pixel contains 40 cells, divided into a simple excitatory (S), complex excitatory (C), and an inhibitory (I) population. The full state $\boldsymbol f=((\boldsymbol f^S)^{\mathsf T},(\boldsymbol f^C)^{\mathsf T},(\boldsymbol f^I)^{\mathsf T})^{\mathsf T}$ contains the firing rates of all populations in every pixel  with $\boldsymbol f^Q \in \mathbb{R}^N, Q\in \{S, C, I\}$, and the combined excitatory response is $\boldsymbol f^E=a^S\boldsymbol f^S+a^C\boldsymbol f^C$, where $(a^S, a^C)\approx(0,7.0.3)$ are the fraction of S and C cells among L4 excitatory cells.

\vspace{0.1in}
\noindent\textbf{The structure of local response function.}
For prescribed input streams, the local response function $\Psi^Q$, $Q\in\{S,C,I\}$, gives the mean firing rate of population $Q$ in each pixel. These responses are computed in advance and stored in the response library. Throughout the manuscript, we reserve $\Psi^Q$ for this local input--response relationship and $\Phi$ for the response map of the entire cortical field \citep{Xiao2026}.

The local response is determined by the stimulus-dependent LGN drive and three cortical input channels: L6 feedback, L4 recurrent excitation, and L4 recurrent inhibition. LGN drive is parameterized by the orientation coordinate $o_p$ and model contrast setting $c$. For population $Q$ at pixel $p$, the inputs and resulting response are
\begin{equation}
\label{local_response_function_explicit}
\begin{aligned}
o_p
&=\cos\!\left[2(\theta-\theta_{\mathrm{aff},p})\right],\\
n_p^{QE}(\boldsymbol f)
&=\sum_{p'=1}^{N}\left[
C_{(p,Q)\leftarrow(p',S)}f_{p'}^S+
C_{(p,Q)\leftarrow(p',C)}f_{p'}^C\right],\\
n_p^{QI}(\boldsymbol f)
&=\sum_{p'=1}^{N}C_{(p,Q)\leftarrow(p',I)}b_I(f_{p'}^I),\\
n_p^{QL6}(\boldsymbol f)
&=\mathcal L_{6,p}^Q[\boldsymbol f^E],\\
\boldsymbol x_p^Q(\boldsymbol f)
&=\left(o_p,c,n_p^{QL6}(\boldsymbol f),n_p^{QE}(\boldsymbol f),n_p^{QI}(\boldsymbol f)\right)^{\mathsf T},\\
f_p^{Q,\mathrm{out}}
&=\Psi^Q\!\left(\boldsymbol x_p^Q(\boldsymbol f)\right).
\end{aligned}
\end{equation}
Here $\theta$ is the stimulus orientation and $\theta_{\mathrm{aff},p}$ is the orientation favored by the
LGN afferent configuration assigned to pixel $p$ (depending on the OD intrinsic to it). Angles in trigonometric expressions are in radians; displayed stimulus angles and decoding errors are in degrees. 
Orientation and contrast vary in the present calculations, following the stimulus setup of Chariker et al.~\citep{Chariker2020}; other stimulus properties follow the selected response library. The coefficient $C_{(p,Q)\leftarrow(p',R)}$ is the mean number of presynaptic cells of type $R$ at pixel $p'$ per target cell of type $Q$ at pixel $p$. Consequently, $n_p^{QE}$, $n_p^{QI}$, and $n_p^{QL6}$ are combined incoming spike rates per target cell, in spikes/s. The function $b_I$ specifies the inherited synaptic-depression $b_I(x)=xs_I(x)$ of inhibitory output before spatial pooling.

The operator $\mathcal L_{6,p}^Q$ describes how the L6 input to each pixel is obtained by spatially averaging the surrounding L4E information. The setup follows the binocular model\citep{Xiao2026},
\begin{equation}
\label{eq:L6_input}
u_p^{(6)}=\sum_{p'\in\mathcal P}k^{(6)}_{p\leftarrow p'}f_{p'}^E,
\qquad
\mathcal L_{6,p}^Q[\boldsymbol f^E]=m^{QL6}R_6(u_p^{(6)}),
\qquad
\sum_{p'}k^{(6)}_{p\leftarrow p'}=1.
\end{equation}
Here $\mathcal P$ is the set of pixels, $k^{(6)}$ is a normalized averaging kernel, $m^{QL6}$ is the mean number of presynaptic L6 cells per type-$Q$ target, and $R_6$ is the effective L6 firing-rate response.

\vspace{0.1in}
\noindent\textbf{A DNN surrogate of the local response function.}
We then approximate each library response function $\Psi^Q$ by a separately parameterized DNN $\widehat{\Psi}^Q$. Training inputs are the five components of $\boldsymbol x$, and targets are the corresponding local mean firing rates of $S$, $C$, and $I$ cells in the response library.

Each DNN has an architecture $5\to96\to96\to96\to1$, with three $\tanh$ hidden layers and a softplus output:
\[
\begin{aligned}
\boldsymbol h_p^{Q,(0)}
&=(\boldsymbol x_p^Q-\boldsymbol\mu^Q)\oslash\boldsymbol s^Q,\\
\boldsymbol h_p^{Q,(k)}
&=\tanh\!\left(W^{Q,(k)}\boldsymbol h_p^{Q,(k-1)}+\boldsymbol b^{Q,(k)}\right),
\qquad k=1,2,3,\\
\widehat{\Psi}^Q(\boldsymbol x_p^Q)
&=\operatorname{softplus}\!\left(W^{Q,(4)}\boldsymbol h_p^{Q,(3)}+b^{Q,(4)}\right).
\end{aligned}
\]
The vectors $\boldsymbol\mu^Q\in\mathbb R^5$ and $\boldsymbol s^Q\in(0,\infty)^5$ contain the training-set feature means and positive normalization scales, respectively; $\oslash$ denotes componentwise division. Both activation functions act componentwise, with $\operatorname{softplus}(z)=\log(1+\exp z)$. The weight dimensions are $W^{Q,(1)}\in\mathbb R^{96\times5}$, $W^{Q,(2)},W^{Q,(3)}\in\mathbb R^{96\times96}$, and $W^{Q,(4)}\in\mathbb R^{1\times96}$. Each population-specific network contains 19,297 trainable parameters.

The training of DNNs combines weighted absolute and relative root-mean-square errors with penalties for low-rate overprediction and violations of the prescribed ordering in the orientation coordinate. Their relative weights are $1,3,0.05,0.10$. Optimization uses AdamW with weight decay $10^{-5}$, batch size 8,192, and learning-rate stages $\{5,2,0.9,0.6,0.2\}\times 10^{-3}$. Target-rate stratification uses boundaries $[0,0.5,2,5,12,22,\infty)$ for S, $[0,2,10,35,65,85,\infty)$ for C, and $[0,1,6,28,73,112,\infty)$ for I.

Model-level matched-rate validation compares 112,000 pixel--condition pairs from 1,600 pixels and 70 sampled stimuli between the CG and DNN cortical calculations (Fig. 1C).

\vspace{0.1in}
\noindent\textbf{Assembling the full-field firing rate response.} For a selected drifting grating, the full response map for the whole field is assembled by evaluating these local functions at the inputs generated by the current state:
\begin{equation}
\label{Phi}
[\Phi(\boldsymbol f)]_p^Q
=\Psi^Q\!\left(o_p,c,n_p^{QL6}(\boldsymbol f),n_p^{QE}(\boldsymbol f),n_p^{QI}(\boldsymbol f)\right),
\qquad Q\in\{S,C,I\},\quad p\in\mathcal P.
\end{equation}
For pixels containing multiple LGN afferent alignments, including pixels near OD boundaries, we use the prescribed output-averaging scheme of the binocular model\citep{Xiao2026} (Sect.~S3B, Eq.~17):
\begin{equation}
\label{eq:OD_mixture}
[\Phi(\boldsymbol f)]_p^Q
=\sum_a w_p^a\Psi^Q\!\left(\cos[2(\theta-a)],c,n_p^{QL6}(\boldsymbol f),n_p^{QE}(\boldsymbol f),n_p^{QI}(\boldsymbol f)\right).
\end{equation}
The index $a$ specifies afferent alignment and $w_p^a$ gives its prescribed output-averaging weight.

\subsection*{Equilibrium computation, benchmarking, and trajectory integration}

For a fixed visual stimulus, we use the full-field response map in Eq.~\ref{Phi}, with the output averaging in Eq.~\ref{eq:OD_mixture} where applicable. In the neural-network surrogate, $\Psi^Q$ is replaced by $\widehat{\Psi}^Q$. 

Equilibria solve $\Phi(\boldsymbol{f})-\boldsymbol{f}=0$. For the benchmark in Fig.~\ref{fig:1}B, CG and surrogate iterations stop when the relative HC-norm step residual reaches $5\times10^{-3}$ (0.5\%). Timing begins after model loading, connectivity construction, and static precomputation, and includes the recurrent solve at orientation $0^\circ$ and contrast setting 100. Field size ranges from $3\times3$ to $40\times40$ hypercolumns, corresponding to $36\times10^3$ to $6.4\times10^6$ neurons in the spiking network model. Repeated timings are summarized by means and standard errors. The separate SNN benchmark focuses on the $3\times3$ hypercolumn case, using 250-s simulations with a 0.1-ms step, 16 CPUs, and 32 GB per task. The compounded order-$10^5$ estimate combines the earlier CG-to-spiking acceleration with the present approximately 100-fold per-update acceleration. Local-library preparation and surrogate training are excluded. 

At an fixed point $\boldsymbol{f^*}$, the damped self-consistency update in Eq.~\eqref{eq:pre-model} has derivative $M_h=(1-h)\mathrm{Id}+hJ$; its local exponential convergence requires $\rho(M_h)<1$, where $\rho$ is the spectral radius. Instead, in our rate model Eq.~\eqref{eq:model}, small perturbations
about an equilibrium evolve according to
$\delta\dot{\boldsymbol f}=L\delta\boldsymbol f$, where
$L=(J-\mathrm{Id})/\tau$ and local exponential stability requires all eigenvalues of $L$ to
have negative real parts. Since $\tau>0$, this is equivalent to
$\alpha(J)<1$, where
$\alpha(J)=\max_j\operatorname{Re}\lambda_j(J)$. Nonlinear trajectories use Eq.~\eqref{eq:model} and \texttt{ode45}. Linear trajectories use $\delta \boldsymbol{f}(t)=e^{Lt}\delta \boldsymbol{f}(0)$. The common observation window is 0--180 ms. Numerical equilibrium-solver iterations have no assigned biological duration.

\subsection*{Calibration and interpretation of model time}

The time constant was obtained by matching the peak time of maximized transient gain to V1 dynamical measurements. To describe this calibration, write dimensionless time as $s=t/\tau$. Equation~\eqref{eq:model} becomes $\mathrm d\boldsymbol{f}/\mathrm ds=\Phi(\boldsymbol{f})-\boldsymbol{f}$. For a fixed reference state, perturbation metric, and readout, changing the common $\tau$ rescales the time axis of the continuous-time gain curve. If that curve reaches its maximum at dimensionless horizon $s_{\mathrm{peak}}$, matching a measured peak time $t_{\mathrm{V1}}$ gives
\begin{equation}
 \tau=\frac{t_{\mathrm{V1}}}{s_{\mathrm{peak}}}.
\end{equation}
For example, at an equilibrium the full-state Euclidean gain curve is $G_\tau(t)=\|\exp[(J-\mathrm{Id})t/\tau]\|_2=G_1(t/\tau)$. The same rescaling applies along a nonlinear trajectory when the gain calculation starts from the same state on that trajectory.
We therefore fix the dimensionless snapshot time $s_0=t_0/\tau$:
changing $\tau$ rescales both the physical snapshot time $t_0$
and the subsequent observation horizon proportionally.
It does not change the equilibrium responses, eigenvectors, stability criterion, and maximum gain over a continuous horizon search. 

We stress the difference between the rate model and the origianl Chariker model series. In this manuscript, the state variables are population mean rates associated with local dynamic equilibria of the driven spiking model. Their time evolution is a closure based on the response functions evaluated at the changing inputs. Unlike the spiking network models of V1\citep{Chariker2016,Chariker2020,chariker2022computational}, fast spike fluctuations and gamma-phase dynamics are averaged out when local equilibria are computed and learned by DNNs. Only their effects on the mean local responses enter the response functions $\Psi$ or $\widehat{\Psi}$.

\subsection*{Response error and perturbation size}

For a firing-rate configuration $\boldsymbol f$, we follow the HC-norm used in the earlier multiscale models\citep{Xiao2024,Xiao2026}:
\begin{equation}
 |\boldsymbol f|_{\mathrm{HC}}=\left\{\frac1N\sum_{p=1}^N
 [0.8(f_p^E)^2+0.2(f_p^I)^2]\right\}^{1/2},
 \qquad \boldsymbol{f}^E=a^S\boldsymbol{f}^S+a^C\boldsymbol{f}^C.
 \label{eq:HC}
\end{equation}
Applied to a displacement $\delta\boldsymbol f$ or a difference between steady responses, it measures a weighted RMS change in the E and I maps, in spikes/s. It is a norm on these two maps and a seminorm on the full S/C/I state, since opposite S and C changes can cancel in E. All full-state singular values and finite-time optimal gains use the Euclidean norm instead. The positive random initialization specified for the perturbation analysis is $\delta \boldsymbol{f}(0)=a|z|$, where $z$ has independent standard Gaussian components and $a$ is chosen so that $|\delta \boldsymbol{f}(0)|_{\mathrm{HC}}=1$ spike/s. The compensation trajectories use one common initial displacement, added to each condition's own equilibrium.

\subsection*{SNN simulation duration}

For the timing comparison, the CG and DNN solvers stop at a relative HC step residual of $\varepsilon=5\times10^{-3}=0.5\%$. To characterize the sampling variability of the SNN firing-rate maps, we performed five independent 250-s simulations, discarded the initial 50 s, and computed a mean E/I rate map from the final 200 s of each run. Let $\bar{\boldsymbol f}_j$ denote the map from run $j$ and $\bar{\boldsymbol f}=\frac15\sum_{j=1}^5\bar{\boldsymbol f}_j$. The sample variance across runs in the HC norm is
\begin{equation}
 V_{\mathrm{HC}}=\frac14\sum_{j=1}^5|\bar{\boldsymbol f}_j-\bar{\boldsymbol f}|_{\mathrm{HC}}^2
 =0.12\,(\mathrm{spikes/s})^2.
\end{equation}
The corresponding relative standard deviation is
\begin{equation}
 \eta_{\mathrm{SNN}}=\frac{\sqrt{V_{\mathrm{HC}}}}{|\bar{\boldsymbol f}|_{\mathrm{HC}}}
 =\frac{0.35}{32.80}\approx1.06\%.
\end{equation}
The 200-s estimate therefore has sampling variability of the same order as the 0.5\% solver residual, motivating the 250-s duration including burn-in. 

\subsection*{Pathway derivatives and fixed-point-preserving interventions}

The chain rule applied to the local response networks and input maps gives $J=J_6+J_E+J_I$. For $a\in\{6,E,I\}$, write $n_p^{Q6}=n_p^{QL6}$ and let $d_p^{Q,a}$ be the local response derivative with respect to that input channel. Then
\begin{equation}
 (J_a)_{(p,Q),(p',R)}
 =d_p^{Q,a}\,\frac{\partial n_p^{Qa}}{\partial f_{p'}^R}
 \bigg|_{\boldsymbol f^*},\qquad
 d_p^{Q,a}=\frac{\partial\widehat\Psi^Q}{\partial n^{Qa}}
 \bigg|_{\text{reference inputs}}.
 \label{eq:pathway_derivatives}
\end{equation}
When the average scheme Eq.~\eqref{eq:OD_mixture} is used, $d_p^{Q,a}$ is the corresponding weighted sum of local derivatives over LGN afferent alignments. Derivatives with respect to the physical input rates include the conversion to library coordinates and feature standardization. The input-map derivatives include nonlinear preprocessing: inhibitory depression contributes $b_I'(f_{p'}^I)$, and the inherited L6 rule contributes the derivative of $R_6$ through the spatial average in Eq.~\eqref{eq:L6_input}. All pathway matrices are evaluated at the same full-circuit equilibrium. 

For a pathway input $a(\boldsymbol f)$, its fixed-point-preserving version is
\begin{equation}
 a_\beta(\boldsymbol f)=a(\boldsymbol f^*)+(1+\beta)[a(\boldsymbol f)-a(\boldsymbol f^*)].
 \label{eq:fpp}
\end{equation}
Same as Sect.~\ref{sec:stability}, $\beta$ denotes the fractional change in the pathway's response to deviations from equilibrium.

Consequently,
\begin{equation}
 J(\beta_6,\beta_I)=J+\beta_6J_6+\beta_IJ_I.
 \label{eq:fpp_Jac}
\end{equation}
On the other hand, when the fixed point is not controlled, the actual strength changes directly multiply the designated pathway drive by $1+\beta_a$ and require finding its new equilibrium. Their derivatives are also evaluated at that equilibrium. 


For a pathway $a\in\{6,E,I\}$ and a matched eigenpair $Jv=\lambda v$, $u^*J=\lambda u^*$, with $u^*v\ne0$, the pathway contribution is defined as: 
\begin{equation}
 c_a=\operatorname{Re}\frac{u^*J_av}{u^*v},\qquad
 -1+c_6+c_E+c_I=\operatorname{Re}\lambda-1.
 \label{eq:budget}
\end{equation}
For a simple eigenvalue this quotient is also the first-order sensitivity to the centered matrix change $J+\epsilon J_a$. The population-relaxation term $-\boldsymbol f/\tau$, labelled ``leak'' in Fig.~\ref{fig:2}B, supplies the $-1$ contribution. The leading eigenvalues of the separate pathway matrices are not summed. The recovery time plotted in Fig.~\ref{fig:7}B is the asymptotic envelope time $\tau/[1-\alpha(J)]$, where $\alpha(J)=\max\operatorname{Re}\lambda(J)<1$.

\subsection*{Spatial kernels and excitatory--inhibitory elimination}

For the Fourier analysis, let $\kappa$ denote the representative E-to-E connectivity kernel normalized to sum to one on the periodic field. This normalization is used for comparing spatial gains; the input calculations retain their prescribed total connection counts. Its gain at spatial frequency $(f_x,f_y)$ is $|\widehat\kappa(f_x,f_y)|/|\widehat\kappa(0,0)|$. The sampled frequencies are $\{-20,\ldots,19\}/40$ cycles/pixel in each direction. The reference gain is 0.05. The flattened, left-half, and noisy kernels in Fig.~\ref{fig:4} preserve prescribed total incoming strength, and their equilibria and Jacobians are recomputed. 

For the cancellation calculation in Supplementary Fig.~S7, write
\begin{equation}
 J=\begin{pmatrix}A&B\\C&D\end{pmatrix},\qquad
 \boldsymbol{f}=\begin{pmatrix}\boldsymbol{f}^{SC}\\\boldsymbol{f}^I\end{pmatrix}, \text{ where}\ \boldsymbol{f}^{SC} = \begin{pmatrix}\boldsymbol{f}^{S}\\\boldsymbol{f}^C\end{pmatrix}.
\end{equation}
The excitatory block $A$ has dimension $3200\times3200$ and retains S and C separately; $D$ has dimension $1600\times1600$. Where $A$ is invertible, eliminating $\boldsymbol{f}^{SC}$ at spectral parameter zero gives $H_0=D-CA^{-1}B$. For each unit eigenvector $v_k$ of $H_0$, the component magnitudes and cancellation residual are
\begin{equation}
 a_k=\|Dv_k\|_2,\qquad b_k=\|CA^{-1}Bv_k\|_2,\qquad
 \eta_k=\frac{\|(D-CA^{-1}B)v_k\|_2}{a_k+b_k}.
\end{equation}
The residual measures relative cancellation of the vectors, including their directions. For the lifted full-state pattern $z_k=(-A^{-1}Bv_k,v_k)$, $Jz_k=(0,H_0v_k)$; its absolute response gain would be $\|H_0v_k\|_2/\|z_k\|_2$. The full eigenvalue equation involves $D-\lambda\mathrm{Id}-C(A-\lambda\mathrm{Id})^{-1}B$. Thus $H_0$ describes the response-map action after eliminating the excitatory variables at spectral parameter zero; the full spectral problem retains the dependence on $\lambda$.

In Fig.~\ref{fig:4}, real-part bins have width 0.1; the blue bin spans $[-0.05,0.05)$. The right column shows the normalized Fourier gains of the representative kernel, rather than singular values of the full Jacobian. Gains are grouped into logarithmically spaced bins and separated at 0.05.

\subsection*{Eigencluster envelopes, singular modes, and finite-time gain}

The leading eigencluster considers the eigenmodes whose eigenvalue real parts lie within $10^{-3}$ of the largest real part, $\alpha=\max_j\operatorname{Re}\lambda_j$, namely $\alpha-\operatorname{Re}\lambda_k\leq10^{-3}$.
Leading right and left eigenvectors are computed separately and their selected spans orthonormalized by rank-checked QR factorization. For an orthonormal basis matrix $V$, partitioned into S, C, and I rows, the E envelope is
\begin{equation}
 e_p=\left[\sum_k|(a^SV_S+a^CV_C)_{pk}|^2\right]^{1/2}.
\end{equation}
Here $p\in\mathcal P$ represents pixel in the field and $k$ indexes the orthonormal
basis vectors spanning the selected eigencluster; the subscript
$pk$ denotes the entry in row $p$ and column $k$. Displayed E envelopes are normalized to unit E-map norm. The example in Fig.~\ref{fig:5}A contains two leading vectors per cluster. These nonnegative envelopes describe spatial support and discard phase and sign. Envelope similarity therefore compares support, rather than principal angles between the invariant subspaces.

Singular vectors satisfy
\begin{equation}
 Jv_1=\sigma_1u_1,\qquad \|v_1\|_2=\|u_1\|_2=1.
 \label{eq:svd}
\end{equation}
For an initial perturbation $\delta \boldsymbol{f}(0)=v_1$, the recurrent component is $Jv_1=\sigma_1u_1$ and the initial rate derivative is $\dot{\delta \boldsymbol{f}}(0)=(\sigma_1u_1-v_1)/\tau$. Thus the singular patterns of $J$ describe the recruited recurrent component of a transient response; finite-time optimal patterns are singular vectors of the propagator defined below.  The vectors are normalized in the full state before population extraction.
Figure~\ref{fig:5}B additionally normalizes each displayed inhibitory component to unit norm, so the maps compare spatial shape.

Similarity uses the sign-insensitive cosine $|a^*b|/(\|a\|_2\|b\|_2)$ on complete S/C/I representations. The main comparisons use 16 combinations of four orientations and four contrast settings. Bars show means and sample standard deviations. For the singular-output controls, 52 conditions cover $0^\circ$ to $90^\circ$ in $7.5^\circ$ steps at four contrasts. For the perturbation control, independent Gaussian noise scaled by each population's mean component value is added to each compared spatial vector, and 500 cosine similarities are computed per condition. Orientation permutations pair vectors at the same contrast but different orientations, producing 528 ordered comparisons separated by at least $15^\circ$ within this set. The upper controls use the full-system/L6 pairing and the lower controls use the full-system/inhibition pairing. The plotted ``Rand.'' reference uses 100,000 independent signed Gaussian vector pairs in 4,800 dimensions. This signed-vector reference supplies a baseline for raw-vector similarity. Nonnegative envelopes require a random reference constructed by applying the same envelope operation to random subspaces of matching rank. The envelope interpretation compares the full--L6 alignment with the other pathway pairs and the orientation-permuted envelope control. The signed-Gaussian reference is not used to assess envelope similarity.

At snapshot time $t$, the finite-time calculation follows the continuing nonlinear trajectory. 
We compute how a small perturbation evolves
over the subsequent interval $[t,t+T]$. The variational propagator $M(t,T)$ maps the perturbation at time
$t$ to its linearized value at time $t+T$: $\delta\boldsymbol f(t+T)=M(t,T)\delta\boldsymbol f(t)$. For durations $T$ that are integer multiples of the time step
$\Delta t=0.5$ ms, we approximate this propagator by
\begin{equation}
 M(t,T)\simeq\prod_{j=0}^{T/\Delta t-1}
 \left[\mathrm{Id}+\frac{\Delta t}{\tau}
 \{J(\boldsymbol{f}(t+j\Delta t))-\mathrm{Id}\}\right],\qquad \Delta t=0.5\ \mathrm{ms}.
\end{equation}
The rightmost matrix acts first on the perturbation, and successive
time steps are applied by multiplying additional matrices on the left. The largest singular value gives the optimal gain at each sampled horizon; maximizing these gains over horizons determines the peak gain and its time. Optimal gains use full-state Euclidean normalization. The pathway-freezing curves are shown in Supplementary Fig.~S8.

\subsection*{Local orientation readout}

The readout in Fig.~\ref{fig:5}E measures small orientation deviations around a reference stimulus; it is not a global decoder over $0$--$180^\circ$. At the reference equilibrium, define $g=\boldsymbol{f}^*$, $q_\theta=\partial_\theta \boldsymbol{f}^*$, and $q_c=\partial_{\log c}\boldsymbol{f}^*$. The response derivatives satisfy $(\mathrm{Id}-J)q_\theta=\partial_\theta\Phi$ and the similar equation for contrast. Let $U$ be an orthonormal basis for the span of $g$ and $q_c$. Then
\begin{equation}
 \widetilde q_\theta=(\mathrm{Id}-UU^*)q_\theta,
 \qquad w_\theta=\frac{\widetilde q_\theta}{\|\widetilde q_\theta\|_2^2},
 \qquad\delta\widehat\theta=w_\theta^*\delta \boldsymbol{f}.
\end{equation}
For $\widetilde q_\theta\ne0$, this gives $w_\theta^*q_\theta=1$ and $w_\theta^*g=w_\theta^*q_c=0$. Orientation units are chosen consistently with the degree scale of the plotted readout. Region masks partition pixels by preferred-orientation distance from the stimulus, modulo $180^\circ$: preferred $d\le15^\circ$, neighboring $15^\circ<d<67.5^\circ$, and orthogonal $d\ge67.5^\circ$. Each region's contribution is $w_\theta^*M_R\delta \boldsymbol{f}$, with the same global normalization and the mask repeated across S/C/I. The contributions sum to the full readout. Each contribution depends jointly on the changing rates and the readout weights, rather than on activity magnitude alone. 

\subsection*{Compensation calculations}


For actual changes to pathway $a\in\{6,E,I\}$, let $\boldsymbol{d}_a=\partial_{\beta_a}\boldsymbol{f}^*$ at baseline. Differentiation of the fixed-point equation gives $(\mathrm{Id}-J)\boldsymbol{d}_a=\partial_{\beta_a}\Phi$. The local least-error direction $\beta_y=m\beta_x$ has slope
\begin{equation}
 m=-\frac{\langle \boldsymbol{d}_y,\boldsymbol{d}_x\rangle_\HC}{\langle \boldsymbol{d}_y,\boldsymbol{d}_y\rangle_\HC},
\end{equation}
provided the denominator is nonzero. The bilinear form is that induced by the E/I response metric in Eq.~\eqref{eq:HC}. The L6--inhibition calculation uses a parameter grid from 0 to 0.4 in steps of 0.008. For each positive $x_j=\beta_6$, the finite sampled minimum in $\beta_I$ gives $y_j$; the through-origin regression slope is $\sum_jx_jy_j/\sum_jx_j^2$. The other parameter planes use the same response-error criterion. The I-to-E comparison in Supplementary Fig.~S11 changes only the I-to-S and I-to-C inputs, whereas whole-inhibition changes also include I-to-I input.

Compensation is fitted to the aggregate E/I map at the reference stimulus; the tuning curves provide a separate comparison across orientations. The fitted line in Figs.~\ref{fig:6}A and \ref{fig:7}B is $\beta_I=0.56\beta_6$. Response errors above 170 spikes/s are excluded from the compensation maps, as indicated by the grey regions.

The three example conditions in Fig.~\ref{fig:7}A use $(\beta_6,\beta_I)=(0,0),(0.072,0.040),(0.296,0.168)$, which lie near the fitted line. Their trajectory distances are measured from their respective equilibria. The corresponding tuning curves are shown in Fig.~\ref{fig:6}B.  



\section*{Acknowledgement}
ZCX is supported by New York University Shanghai through boost fund and the Shanghai Municipal Education Commission through talent programs. ZB is supported by the Engineering and Physical Sciences Research Council (EPSRC, UK) [grant number EP/Z533658/1] and the São Paulo Research Foundation (FAPESP, Brazil) [grant number 2023/13706-0]. During the time of the work, XH was supported by New York University Shanghai through Deans' Undergraduate Research Fund (DURF), and QY was supported by New York University Shanghai through Summer Research Program in Neuroscience (SRPN). 

We thank Professor Lai-Sang Young (New York University) and Professor Kevin K Lin (University of Arizona) for useful discussions.
\section*{Contribution}
The project is conceptualized and supervised by ZCX and ZB. XH takes care of software, code writing, and figures. XH, ZCX, and ZB analyze the data, performed theoretical analysis, and write the manuscript. QY contributes to the artificial neural network training for an earlier version of the model (not included in the manuscript).

\section*{Code and Data Availability}
The codes and Data of this project is available at: \url{https://github.com/XialeiH/Spectral\_Analysis\_V1}.

\bibliographystyle{naturemag}
\bibliography{manuscript-1017}

@article{BenYishai1995,
  author = {Ben-Yishai, R. and Bar-Or, R. L. and Sompolinsky, H.},
  title = {{Theory of orientation tuning in visual cortex}},
  journal = {Proc. Natl Acad. Sci. USA},
  year = {1995},
  volume = {92},
  pages = {3844--3848},
  doi = {10.1073/pnas.92.9.3844}
}

@article{Prinz2004,
  author = {Prinz, A. A. and Bucher, D. and Marder, E.},
  title = {{Similar network activity from disparate circuit parameters}},
  journal = {Nat. Neurosci.},
  year = {2004},
  volume = {7},
  pages = {1345--1352},
  doi = {10.1038/nn1352}
}

@article{Powell2021,
  author = {Powell, D. J. and Marder, E. and Nusbaum, M. P.},
  title = {{Perturbation-specific responses by two neural circuits generating similar activity patterns}},
  journal = {Curr. Biol.},
  year = {2021},
  volume = {31},
  pages = {4831--4838.e4},
  doi = {10.1016/j.cub.2021.08.042}
}

@article{Chariker2016,
  author = {Chariker, L. and Shapley, R. and Young, L.-S.},
  title = {{Orientation selectivity from very sparse LGN inputs in a comprehensive model of macaque V1 cortex}},
  journal = {J. Neurosci.},
  year = {2016},
  volume = {36},
  pages = {12368--12384},
  doi = {10.1523/JNEUROSCI.2603-16.2016}
}

@article{Chariker2020,
  author = {Chariker, L. and Shapley, R. and Young, L.-S.},
  title = {{Contrast response in a comprehensive network model of macaque V1}},
  journal = {J. Vis.},
  year = {2020},
  volume = {20},
  number = {4},
  pages = {16},
  doi = {10.1167/jov.20.4.16}
}

@unpublished{Xiao2026,
  author = {Xiao, Z.-C. and Lin, K. K. and Huang, X. and Young, L.-S.},
  title = {{Biology-constrained multiscale models for inferring neuroanatomy, with application to primate visual cortex}},
  year = {2026},
  note = {arXiv preprint, arXiv:2606.21785}
}

@article{Xiao2024,
  author = {Xiao, Z.-C. and Lin, K. K. and Young, L.-S.},
  title = {{Efficient models of cortical activity via local dynamic equilibria and coarse-grained interactions}},
  journal = {Proc. Natl Acad. Sci. USA},
  year = {2024},
  volume = {121},
  pages = {e2320454121},
  doi = {10.1073/pnas.2320454121}
}

@article{Yarotsky2017,
  author = {Yarotsky, D.},
  title = {{Error bounds for approximations with deep ReLU networks}},
  journal = {Neural Netw.},
  year = {2017},
  volume = {94},
  pages = {103--114},
  doi = {10.1016/j.neunet.2017.07.002}
}

@article{Poggio2017,
  author = {Poggio, T. and Mhaskar, H. and Rosasco, L. and Miranda, B. and Liao, Q.},
  title = {{Why and when can deep---but not shallow---networks avoid the curse of dimensionality: a review}},
  journal = {Int. J. Autom. Comput.},
  year = {2017},
  volume = {14},
  pages = {503--519},
  doi = {10.1007/s11633-017-1054-2}
}

@article{Greenbaum2020,
  author = {Greenbaum, A. and Li, R.-C. and Overton, M. L.},
  title = {{First-order perturbation theory for eigenvalues and eigenvectors}},
  journal = {SIAM Rev.},
  year = {2020},
  volume = {62},
  pages = {463--482},
  doi = {10.1137/19M124784X}
}

@article{Ozeki2009,
  author = {Ozeki, H. and Finn, I. M. and Schaffer, E. S. and Miller, K. D. and Ferster, D.},
  title = {{Inhibitory stabilization of the cortical network underlies visual surround suppression}},
  journal = {Neuron},
  year = {2009},
  volume = {62},
  pages = {578--592},
  doi = {10.1016/j.neuron.2009.03.028}
}

@article{Rubin2015,
  author = {Rubin, D. B. and Van Hooser, S. D. and Miller, K. D.},
  title = {{The stabilized supralinear network: a unifying circuit motif underlying multi-input integration in sensory cortex}},
  journal = {Neuron},
  year = {2015},
  volume = {85},
  pages = {402--417},
  doi = {10.1016/j.neuron.2014.12.026}
}

@article{Mastrogiuseppe2018,
  author = {Mastrogiuseppe, F. and Ostojic, S.},
  title = {{Linking connectivity, dynamics, and computations in low-rank recurrent neural networks}},
  journal = {Neuron},
  year = {2018},
  volume = {99},
  pages = {609--623.e29},
  doi = {10.1016/j.neuron.2018.07.003}
}

@article{MurphyMiller2009,
  author = {Murphy, B. K. and Miller, K. D.},
  title = {{Balanced amplification: a new mechanism of selective amplification of neural activity patterns}},
  journal = {Neuron},
  year = {2009},
  volume = {61},
  pages = {635--648},
  doi = {10.1016/j.neuron.2009.02.005}
}

@article{Hennequin2012,
  author = {Hennequin, G. and Vogels, T. P. and Gerstner, W.},
  title = {{Non-normal amplification in random balanced neuronal networks}},
  journal = {Phys. Rev. E},
  year = {2012},
  volume = {86},
  pages = {011909},
  doi = {10.1103/PhysRevE.86.011909}
}

@article{Cadena2019,
  author = {Cadena, S. A. and Denfield, G. H. and Walker, E. Y. and Gatys, L. A. and Tolias, A. S. and Bethge, M. and Ecker, A. S.},
  title = {{Deep convolutional models improve predictions of macaque V1 responses to natural images}},
  journal = {PLoS Comput. Biol.},
  year = {2019},
  volume = {15},
  pages = {e1006897},
  doi = {10.1371/journal.pcbi.1006897}
}

@article{Wang2025,
  author = {Wang, E. Y. and others},
  title = {{Foundation model of neural activity predicts response to new stimulus types}},
  journal = {Nature},
  year = {2025},
  volume = {640},
  pages = {470--477},
  doi = {10.1038/s41586-025-08829-y}
}

@misc{Zhang2026Diffusion,
  author = {Zhang, Ruilin and Tao, Louis and Xiao, Zhuo-Cheng},
  title = {Diffusion learning reveals viable parameter manifolds and compensation geometry in biological dynamical systems},
  year = {2026},
  eprint = {2607.03671},
  archivePrefix = {arXiv},
  primaryClass = {q-bio.QM},
  doi = {10.48550/arXiv.2607.03671},
  url = {https://arxiv.org/abs/2607.03671},
  note = {Version 2, September 11, 2026}
}

@article{Goncalves2020,
  author = {Gon{\c{c}}alves, Pedro J. and Lueckmann, Jan-Matthis and Deistler, Michael and Nonnenmacher, Marcel and {\"O}cal, Kaan and Bassetto, Giacomo and Chintaluri, Chaitanya and Podlaski, William F. and Haddad, Sara A. and Vogels, Tim P. and Greenberg, David S. and Macke, Jakob H.},
  title = {Training deep neural density estimators to identify mechanistic models of neural dynamics},
  journal = {eLife},
  year = {2020},
  volume = {9},
  pages = {e56261},
  doi = {10.7554/eLife.56261},
  url = {https://elifesciences.org/articles/56261}
}

@article{Reinhold2015,
  author = {Reinhold, Kimberly and Lien, Anthony D. and Scanziani, Massimo},
  title = {Distinct recurrent versus afferent dynamics in cortical visual processing},
  journal = {Nat. Neurosci.},
  year = {2015},
  volume = {18},
  number = {12},
  pages = {1789--1797},
  doi = {10.1038/nn.4153}
}

@article{Webster1990,
  author = {Webster, Michael A. and {De Valois}, Karen K. and Switkes, Eugene},
  title = {Orientation and spatial-frequency discrimination for luminance and chromatic gratings},
  journal = {J. Opt. Soc. Am. A},
  year = {1990},
  volume = {7},
  number = {6},
  pages = {1034--1049},
  doi = {10.1364/JOSAA.7.001034}
}

@article{Moore2005,
  author = {Moore, IV, Bartlett D. and Alitto, Henry J. and Usrey, W. Martin},
  title = {Orientation tuning, but not direction selectivity, is invariant to temporal frequency in primary visual cortex},
  journal = {J. Neurophysiol.},
  year = {2005},
  volume = {94},
  number = {2},
  pages = {1336--1345},
  doi = {10.1152/jn.01224.2004}
}

@article{Priebe2006,
  author = {Priebe, Nicholas J. and Lisberger, Stephen G. and Movshon, J. Anthony},
  title = {Tuning for Spatiotemporal Frequency and Speed in Directionally Selective Neurons of Macaque Striate Cortex},
  journal = {J. Neurosci.},
  year = {2006},
  volume = {26},
  number = {11},
  pages = {2941--2950},
  doi = {10.1523/JNEUROSCI.3936-05.2006}
}

@article{Ringach2002,
  author = {Ringach, D. L. and Shapley, R. M. and Hawken, M. J.},
  title = {{Orientation selectivity in macaque V1: diversity and laminar dependence}},
  journal = {J. Neurosci.},
  year = {2002},
  volume = {22},
  number = {13},
  pages = {5639--5651},
  doi = {10.1523/JNEUROSCI.22-13-05639.2002}
}

@article{ConnollyVanEssen1984,
  author = {Connolly, M. and {Van Essen}, D.},
  title = {{The representation of the visual field in parvicellular and magnocellular layers of the lateral geniculate nucleus in the macaque monkey}},
  journal = {J. Comp. Neurol.},
  year = {1984},
  volume = {226},
  number = {4},
  pages = {544--564},
  doi = {10.1002/cne.902260408}
}

@article{Churchland2010,author = {Churchland, Mark M. and Yu, Byron M. and Cunningham, John P. and Sugrue, Leo P. and Cohen, Marlene R. and Corrado, Greg S. and Newsome, William T. and Clark, Andrew M. and Hosseini, Paymon and Scott, Benjamin B. and Bradley, David C. and Smith, Matthew A. and Kohn, Adam and Movshon, J. Anthony and Armstrong, Katherine M. and Moore, Tirin and Chang, Steve W. and Snyder, Lawrence H. and Lisberger, Stephen G. and Priebe, Nicholas J. and Finn, Ian M. and Ferster, David and Ryu, Stephen I. and Santhanam, Gopal and Sahani, Maneesh and Shenoy, Krishna V.},
  title = {{Stimulus onset quenches neural variability: a widespread cortical phenomenon}},
  journal = {Nat. Neurosci.},
  year = {2010},
  volume = {13},
  number = {3},
  pages = {369--378},
  doi = {10.1038/nn.2501}
}

@article{Olsen2012,
  author = {Olsen, Shawn R. and Bortone, Dante S. and Adesnik, Hillel and Scanziani, Massimo},
  title = {Gain control by layer six in cortical circuits of vision},
  journal = {Nature},
  year = {2012},
  volume = {483},
  number = {7387},
  pages = {47--52},
  doi = {10.1038/nature10835},
  url = {https://www.nature.com/articles/nature10835}
}

@article{Bortone2014,
  author = {Bortone, Dante S. and Olsen, Shawn R. and Scanziani, Massimo},
  title = {Translaminar inhibitory cells recruited by layer 6 corticothalamic neurons suppress visual cortex},
  journal = {Neuron},
  year = {2014},
  volume = {82},
  number = {2},
  pages = {474--485},
  doi = {10.1016/j.neuron.2014.02.021},
  url = {https://doi.org/10.1016/j.neuron.2014.02.021}
}

@article{chariker2022computational,
  author = {Chariker, Logan and Shapley, Robert and Hawken, Michael and Young, Lai-Sang},
  title = {A Computational Model of Direction Selectivity in Macaque {V1} Cortex Based on Dynamic Differences between {On} and {Off} Pathways},
  journal = {J. Neurosci.},
  year = {2022},
  volume = {42},
  number = {16},
  pages = {3365--3380},
  doi = {10.1523/JNEUROSCI.2145-21.2022},
  url = {https://www.jneurosci.org/content/42/16/3365}
}

@article{BlasdelFitzpatrick1984,
  author = {Blasdel, Gary G. and Fitzpatrick, David},
  title = {Physiological organization of layer 4 in macaque striate cortex},
  journal = {J. Neurosci.},
  year = {1984},
  volume = {4},
  number = {3},
  pages = {880--895},
  doi = {10.1523/JNEUROSCI.04-03-00880.1984}
}

@article{Hawken1988,
  author = {Hawken, Michael J. and Parker, Andrew J. and Lund, Jennifer S.},
  title = {Laminar organization and contrast sensitivity of direction-selective cells in the striate cortex of the Old World monkey},
  journal = {J. Neurosci.},
  year = {1988},
  volume = {8},
  number = {10},
  pages = {3541--3548},
  doi = {10.1523/JNEUROSCI.08-10-03541.1988}
}

@article{Fitzpatrick1985,
  author = {Fitzpatrick, David and Lund, Jennifer S. and Blasdel, Gary G.},
  title = {Intrinsic connections of macaque striate cortex: afferent and efferent connections of lamina {4C}},
  journal = {J. Neurosci.},
  year = {1985},
  volume = {5},
  number = {12},
  pages = {3329--3349},
  doi = {10.1523/JNEUROSCI.05-12-03329.1985}
}

@article{ObermayerBlasdel1993,
  author = {Obermayer, Klaus and Blasdel, Gary G.},
  title = {Geometry of orientation and ocular dominance columns in monkey striate cortex},
  journal = {J. Neurosci.},
  year = {1993},
  volume = {13},
  number = {10},
  pages = {4114--4129},
  doi = {10.1523/JNEUROSCI.13-10-04114.1993}
}

@article{DerringtonLennie1984,
  author = {Derrington, Andrew M. and Lennie, Peter},
  title = {Spatial and temporal contrast sensitivities of neurones in lateral geniculate nucleus of macaque},
  journal = {J. Physiol.},
  year = {1984},
  volume = {357},
  pages = {219--240},
  doi = {10.1113/jphysiol.1984.sp015498}
}

@article{CasteloBranco1998,
  author = {Castelo-Branco, Miguel and Neuenschwander, Sergio and Singer, Wolf},
  title = {Synchronization of visual responses between the cortex, lateral geniculate nucleus, and retina in the anesthetized cat},
  journal = {J. Neurosci.},
  volume = {18},
  number = {16},
  pages = {6395--6410},
  year = {1998},
  doi = {10.1523/JNEUROSCI.18-16-06395.1998}
}

@article{Han2021,
  author = {Han, Chuanliang and Wang, Tian and Yang, Yi and Wu, Yujie and Li, Yang and Dai, Weifeng and Zhang, Yange and Wang, Bin and Yang, Guanzhong and Cao, Ziqi and Kang, Jian and Wang, Gang and Li, Liang and Yu, Hongbo and Yeh, Chun-I and Xing, Dajun},
  title = {Multiple gamma rhythms carry distinct spatial frequency information in primary visual cortex},
  journal = {PLoS Biol.},
  volume = {19},
  number = {12},
  pages = {e3001466},
  year = {2021},
  doi = {10.1371/journal.pbio.3001466}
}

@article{Bastos2014,
  author = {Bastos, Andre M. and Briggs, Farran and Alitto, Henry J. and Mangun, George R. and Usrey, W. Martin},
  title = {Simultaneous recordings from the primary visual cortex and lateral geniculate nucleus reveal rhythmic interactions and a cortical source for gamma-band oscillations},
  journal = {J. Neurosci.},
  volume = {34},
  number = {22},
  pages = {7639--7644},
  year = {2014},
  doi = {10.1523/JNEUROSCI.4216-13.2014}
}

@article{Trepka2022,
  author = {Trepka, Ethan B. and Zhu, Shude and Xia, Ruobing and Chen, Xiaomo and Moore, Tirin},
  title = {Functional interactions among neurons within single columns of macaque {V1}},
  journal = {eLife},
  volume = {11},
  pages = {e79322},
  year = {2022},
  doi = {10.7554/eLife.79322}
}

@article{Tang2018,
  author = {Tang, Shiming and Zhang, Yimeng and Li, Zhihao and Li, Ming and Liu, Fang and Jiang, Hongfei and Lee, Tai Sing},
  title = {Large-scale two-photon imaging revealed super-sparse population codes in the {V1} superficial layer of awake monkeys},
  journal = {eLife},
  volume = {7},
  pages = {e33370},
  year = {2018},
  doi = {10.7554/eLife.33370}
}

@article{De2020,
  author = {De, Abhishek and El-Shamayleh, Yasmine and Horwitz, Gregory D.},
  title = {Fast and reversible neural inactivation in macaque cortex by optogenetic stimulation of {GABAergic} neurons},
  journal = {eLife},
  volume = {9},
  pages = {e52658},
  year = {2020},
  doi = {10.7554/eLife.52658}
}

@article{xiao2021data,
  author = {Xiao, Zhuo-Cheng and Lin, Kevin K. and Young, Lai-Sang},
  title = {{A data-informed mean-field approach to mapping of cortical parameter landscapes}},
  journal = {PLoS Comput. Biol.},
  volume = {17},
  number = {12},
  pages = {e1009718},
  year = {2021},
  doi = {10.1371/journal.pcbi.1009718}
}

\clearpage

\clearpage

\clearpage

\clearpage

\clearpage

\clearpage

\clearpage
\end{document}